\documentclass[useAMS,usenatbib,twocolumn]{mnras}
\usepackage{floatrow}
\usepackage{caption}
\usepackage{calc}
\usepackage{graphicx}
\usepackage{amsmath}
\usepackage{caption}

\usepackage[utf8]{inputenc}
\usepackage{graphicx}
\usepackage{color}
\usepackage{booktabs}
\usepackage{amssymb}
\usepackage[]{natbib}

\def\apj{ApJ}
\def\apjs{ApJS}
\def\apjl{ApJ}
\def\aj{AJ}
\def\mnras{MNRAS}

\def\nat{Nature}
\def\araa{ARA\&A}
\def\aap{A\&A}

\newcommand{\eagle}{{\sc eagle}}

\newcommand{\cddf}{{\sc cddf}}
\newcommand{\collosus}{{\sc collosus}}
\newcommand{\dla}{{\sc dla}}
\newcommand{\lls}{{\sc lls}}
\newcommand{\col}{{\sc colossus}}
\begin{document}
	
	\title[The \dla-halo connection]{Connecting cosmological accretion to strong Ly$\alpha$ absorbers}

	\author[Tom Theuns] 
	{Tom Theuns$^{\href{https://orcid.org/0000-0002-3790-9520}{\includegraphics[width=0.4cm]{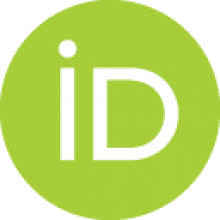}}}$\\
		Institute for Computational Cosmology, Department of Physics, Durham University, DH1 3LE Durham, UK} 
	\date{Accepted --. Received --; in original form --}
	
	\maketitle
	
	\begin{abstract}
		We present an analytical model for the cosmological accretion of gas onto dark matter halos, based on a similarity solution applicable to spherical systems. Performing simplified radiative transfer, we compute how the accreting gas turns increasingly neutral as it self-shields from the ionising background, and obtain the column density, $N_{\rm HI}$,  as a function of impact parameter. The resulting column-density distribution function (\cddf) is in excellent agreement with observations. The analytical expression elucidates (1) why halos over a large range in mass contribute about equally to the \cddf\ as well as (2) why the \cddf\ evolves so little with redshift in the range $z=2\rightarrow 5$. We show that  the model also predicts reasonable \dla\ line-widths ($v_{90}$), bias and molecular fractions. Integrating over the \cddf\ yields the mass density in neutral gas, $\Omega_{\rm HI}$, which agrees well with observations. $\Omega_{\rm HI}(z)$ is nearly constant even though the accretion rate onto halos evolves. We show that this occurs because the fraction of time that the inflowing gas is neutral depends on the dynamical time of the halo, which is inversely proportional to the accretion rate. Encapsulating results from cosmological simulations, the simple model shows that most Lyman-limit system and damped Lyman-$\alpha$ absorbers are associated with the cosmological accretion of gas onto halos.
	\end{abstract}
\section{Introduction}
UV-photons emitted by a cosmologically-distant source, such as for example a quasar, may be absorbed {\em en route} by scattering off neutral Hydrogen atoms in the Lyman ($n=1\rightarrow n'$)  or Lyman-limit ($n=1\rightarrow\infty$) transitions ($n$ and $n'$ are the Hydrogen atom's principle quantum number). This generates a series of absorption lines due to intervening \lq Lyman-$\alpha$\rq\ clouds. The spectral signature when many lines are close together and overlap is that of a \lq forest\rq\ of lines, hence the name \lq Lyman-$\alpha$ forest\rq\ \citep{Weymann81}. The taxonomy of these lines depends on how column-density, $N_{\rm HI}$, is determined.  When $N_{\rm HI}$ is sufficiently small, absorption lines have a nearly Gaussian shape which can be fit accurately with a Voigt profile. Lines become increasingly square in shape and insensitive to $N_{\rm HI}$ with increasing $N_{\rm HI}$, but once $N_{\rm HI}\gtrsim 10^{17.2}{\rm cm}^{-2}$, the Lyman-limit optical depth is so high that the presence of the absorber results in a strong depression of the flux below $(1+z)\times 912\AA$, where $z$ is the absorber's redshift.  Finally at even higher values of $N_{\rm HI}\ge 10^{20.3}{\rm cm}^{-2}$, absorption is so strong that the line-shape can be measured far from the line centre where it is dominated by the natural line-width of the Lyman-$\alpha$ transition. The Lorenzian shape of this natural line-profile leads to such high $N_{\rm HI}$ absorbers to be called \lq damped Lyman-$\alpha$ absorbers\rq\ (\dla s), in analogy with the Lorentzian shape of the resonance of a damped harmonic oscillator \citep{Beaver72, Wolfe86}. Lines are therefore classified based on column-density as Lyman-$\alpha$ forest, Lyman-limit system (\lls s), and \dla s for $N_{\rm HI}<10^{17.2}$, $<10^{20.3}$, and $\ge 10^{20.3}{\rm cm}^{-2}$, respectively \citep[see e.g.][for  reviews]{Rauch98, Peroux20}. This paper focuses on \dla s, foraying briefly into the realm of \lls s.

The original motivation for studying \dla s by \cite{Wolfe86} was to identify high-redshift {\em galaxies}. Indeed, a sight-line through the Milky Way's disk is likely to have $N_{\rm HI}\ge 10^{20.3}{\rm cm}^{-2}$, therefore a sight-line through a high-$z$ Milky Way-like galaxy would likely result in a \dla\ in the spectrum of a background quasar - and, of course, high-redshift \dla s were already known and detectable back then. This realisation inadvertently initiated decades of relatively unsuccessful attempts to identify optical counterpart(s) to \dla s (with some exceptions, e.g. \citealt[][]{Moller04, Fynbo10, Peroux12,Fumagalli15}). The low success rate of such searches was attributed to the \dla\ galaxies being generally too faint to be detectable \cite[see e.g. the discussion by][]{Krogager17}.  This endeavour has been revolutionised by the advent of integral field units ({\sc ifu}'s) which allow for the  identification of faint galaxies at a range of distances from the sight-line to the quasar in a single telescope pointing (e.g. \citealt{Peroux16} using {\sc sinfoni}, \citealt{Fumagalli17, Mackenzie19} using {\sc muse}).  Such studies suffer from the opposite problem: there may be be several galaxies detected close to the sight-line at similar redshift as the \dla, in which case it becomes unclear which particular galaxy - if any - to associate with the \dla. 

The prevailing view that \dla s are gas closely associated with a galaxy was also the motivation for several theoretical models \cite[e.g.][]{Maller01, Fynbo08,DiGioia20,Krogager20}. Typically, these struggle to reproduce the observed number density of \dla s for a realistic value of {\em size} of the galaxy. In the relatively few cases where a galaxy is identified with a particular \dla, it is not unusual for the impact parameter to be $b>10~{\rm kpc}$ (and sometimes much larger than that, e.g. \citealt[][]{Fumagalli17}), and that is of course (much) larger than the sizes of (stellar components of) high-$z$ galaxies, which are closer to $\sim 3~{\rm kpc}$ even for massive galaxies \cite[e.g.][]{VanDerWel14}. 

\dla s are relatively rare, with of order $\sim 0.3$ \dla\ per decade in column-density per redshift interval $\Delta z=1$ at $z=3$. Statistically robust studies of \dla s, especially at high $N_{\rm HI}$, have been revolutionized by the advent of very large samples of quasar spectra (see for example \dla s identified by \citealt{Noterdaeme12} in the {\em Sloan Digital Sky Survey} ({\sc sdss}) data release 9, \citealt{DR9}, or by \citealt{Ho20} in {\sc sdss} DR12, \citealt{SDSS12}).
Whereas earlier numerical simulations \cite[e.g.][]{Nagamine04, Tescari09} and cosmological zoom simulations \cite[e.g.][]{Pontzen08} reproduced the earlier data by \cite{Peroux01, Prochaska05} relatively well, the leap in data brought about by the {\sc sdss} made it harder for simulations to reproduce the data.

Nevertheless, several relatively recent simulations can reproduce the vastly improved \dla\ statistics reasonably well \citep[e.g.][]{Altay11, Cen12,Bird14}. The simulations differ in numerical technique (particle based, grid based, moving mesh, respectively) and vary the prescriptions associated with the formation of the galaxy (both how stars form and importantly in the way that feedback is implemented.) What these simulations have in common is that the majority of the \dla\ cross section is not associated with the stellar component of the galaxy, but rather the gas that gives rise to the \dla\ is distributed throughout and even outside the galaxy's {\em halo},  (\lq every galaxy is a \dla, but not every \dla\ is a galaxy\rq), typically in the form of several filaments of high density gas, as also seen in the high-resolution zoom simulations of \cite{Fumagalli11, Faucher11}. 
In the simulations presented by \cite{VandeVoort12}, the majority of the gas associated with \lls s and \dla s 
at redshift $z\sim 3$ is falling rapidly towards the central galaxy of a dark matter halo while remaining cool (temperature $T \lessapprox  10^{5.5}{\rm K}$). As it accretes onto the galaxy, the majority of gas is ejected in the form of a galactic outflow powered by supernovae. Gas inside the galaxy that is neither inflowing nor outflowing and the outflow itself adds a small contribution to the total \dla\ cross section. 

Two aspects of these simulations form the motivation for the analytical model described here. Firstly, the \dla\ gas is mostly accreting towards the centre of dark matter halos \citep{VandeVoort12}, therefore our model is based on cosmological accretion (rather than assuming \dla\ gas is in centrifugal or hydrostatic equilibrium
in a halo as in the model by \citealt{Padmanabhan16}, or is associated directly with galaxies, as in
the recent models by \citealt{Krogager20, DiGioia20}). We will assume that the accretion is mostly smooth (as opposed to accretion via satellites, say), an idea supported by simulations \citep[e.g.][]{Crain17, Wright20}, at least at higher $z\gtrapprox 1$, say. Both assumptions are consistent with the \lq cold accretion\rq\ paradigm of \cite{Keres05,Dekel09}. Secondly, although the importance of feedback from the galaxy on \dla s is somewhat contested \cite[e.g.][]{Sommer-Larsen17, Rhodin19}, the simulations by \cite{Altay13a} that are based on the {\sc owls} project described by \cite{Schaye10}, include a very wide variety of feedback implementations and show convincingly that outflows driven by feedback affect \dla\ statistics mostly at the high column-density end, $N_{\rm HI}\gtrapprox 10^{21.5}{\rm cm}^{-2}$. This encourages us to simply neglect feedback, either from outflows, or indeed from local ionising radiation which may play at role at the high column-density end \cite[e.g.][]{Rahmati13}. Obviously, this is only an approximation. As stressed by e.g. \citealt[][]{Krogager17}, correlations between \dla s and galaxies depend on metallicity: since we neglect feedback, we cannot study this interesting observational finding. A third motivation for developing an analytical model is to elucidate why the \dla\ column-density distribution ({\sc cddf}) evolves so weakly with redshift (as observed and reproduced by simulations) and why halos with a range of masses contribute about equally to the \cddf. Having an analytical expression for the \cddf\ makes this straightforward. 

This paper is organised as follow: the model is described in section \ref{sect:model}, which starts by discussing the radial distribution of gas accreting onto halos and goes on to derive the resulting \cddf. Section \ref{sect:discussion} discusses in more detail the validity of some of the assumptions and where improvements could be made. Finally,  section \ref{sect:conclusions} summarizes. We will use the \cite{Planck14} values of the cosmological parameters were needed.

\section{The model}
\label{sect:model}
Our model for connecting \dla s to dark matter halos has several ingredients and makes many simplifying assumptions. It is described in \S \ref{sect:accretion}, which starts by relating the radial and column-density distribution of neutral gas to a halo of a given mass at a given $z$, and explores the differential contribution of halos of a given mass to the \cddf. We explain there why a relatively extended range in halos masses contributes significantly to the \cddf. Next, we calculate the \cddf\ in \S \ref{sect:redshift} and use the analytical expression to explore its redshift evolution. The dynamics of the gas is examined in terms of \dla\ bias in \S \ref{sect:bias} and in terms of line-widths in \S \ref{sect:v90}. We explore the molecular contents of \dla s in \S \ref{sect:molecules}, and the cosmological fraction of gas in \dla s in  \S \ref{sect:OmegaHI}.

\subsection{Cosmological accretion onto halos}
\label{sect:accretion}
\begin{figure*}
	\includegraphics[width=.95\textwidth]{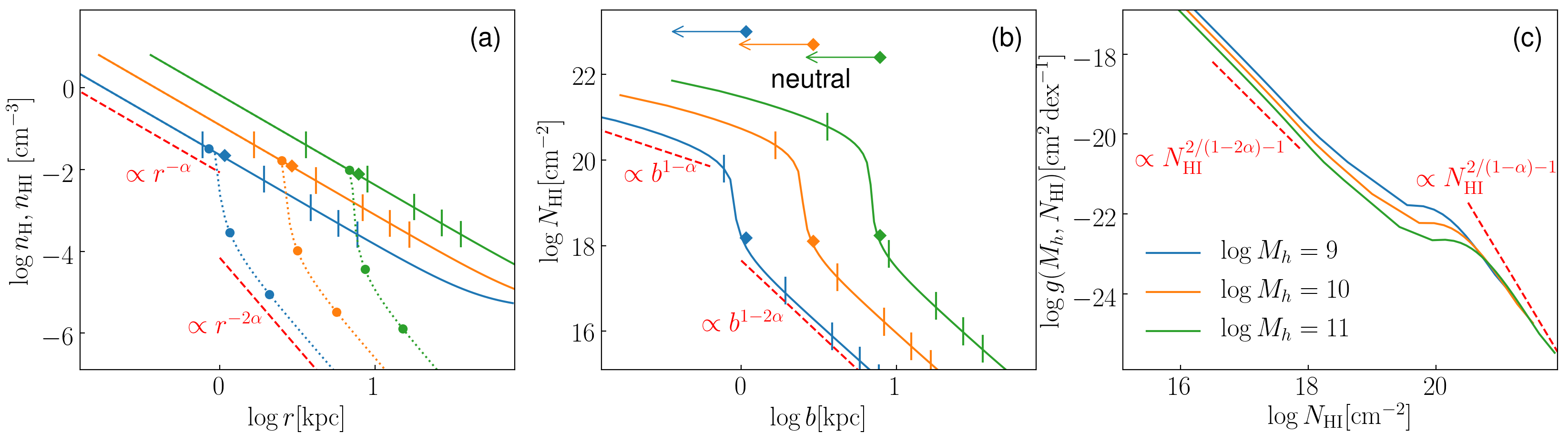}\hfill
	\caption{Gas profiles of dark matter halos at redshift $z=3$ for $\alpha=2.2$. 
		{\bf Panel (a)}: Radial profiles of the total hydrogen density $n_{\rm H}$ ({\em solid lines}) and neutral hydrogen density $n_{\rm HI}$ ({\em dotted lines}) computed by integrating Eq.~(\ref{eq:xneut}) numerically, for different halo masses (different colours as per the legend in panel (c)). {\em Thin vertical lines} are drawn at fractions 0.1, 0.25, 0.5, 0.75 and 1 of the viral radius $R_h$ of the halo; {\em solid dots} are drawn at locations where the optical depth reaches $\tau=10$, 1 and 0.1 (from the inside out); the {\em filled diamond} is drawn at the approximate location of the ionization front $r_I$ from Eq.~(\ref{eq:rI}). Power-law profiles are drawn as {\em red-dashed} lines.		
		{\bf Panel (b)}: Corresponding column-density, $N_{\rm HI}$, as a function of impact parameter $b$. {\em Thin vertical lines} are drawn where $b$ equals a fraction of 0.1, 0.25, 0.5, 0.75 and 1 of the viral radius $R_h$ of the halo; red-dashed inner and outer lines indicate power-laws, $N_{\rm HI}\propto b^{1-\alpha}$ and $N_{\rm HI}\propto b^{1-2\alpha}$; {\em filled diamonds} are drawn at the location of $r_I$, with additionally arrows pointing inwards from where the gas is mostly neutral. {\bf Panel (c):}  Contribution of such halos to the column-density distribution from Eq.~(\ref{eq:gCDDF}). The red-dashed inner and outer lines indicate power-laws with the slopes expected from the dependence of the cross section on column-density (see text). In all three panels, profiles are coloured by the logarithm of the halo mass, $M_h$, in solar masses, as per the legend in panel (c).}
	\label{fig:radial}
\end{figure*}
The equations describing cosmological accretion onto a spherically symmetric over-density associated with a dark matter halo admits a similarity solution in the case of an Einstein-de Sitter (EdS) universe\footnote{For studying structure formation at redshift $z\gtrapprox 1$, an EdS model should be sufficiently accurate.} \citep{Bertschinger85}. In this solution, the density profile is a power-law, $\rho(r)\propto r^{-\alpha}$, over a large range in radii $r$. The value of the exponent, $\alpha$, depends on whether the accreting material is collisional or collisionless and on the inner boundary condition and is in the range $\alpha\sim 1.5-2.2$. We will assume here that accreting gas remains cool while accreting (in fact, we will assume it remains isothermal), and as it enters the interstellar medium of the galaxy, it cools rapidly so that the accretion shock does not propagate outwards. Under these assumptions, $\alpha$ is at the steeper end of the range: we will pick $\alpha=2.2$ for making figures. Hydrodynamical simulations show that the assumptions on the thermal evolution of the accreting gas are reasonable, at least for sufficiently low-mass halos; see for example the density-temperature diagram of gas at over-densities $\rho/\langle\rho\rangle$ in the range 10-100 in Figure~2 of \cite{Theuns98}.

We set the normalisation of the density profile by requiring that the mean density within
the halo's virial radius, $R_h$, is 200 times the critical density \cite[e.g.][]{Mo98}. To calculate the corresponding density profile of hydrogen gas\footnote{We will use a subscript $h$ to denote properties of a halo, and subscript H to denote properties that refer to hydrogen.}, we assume that the  accreting mass has its cosmological share of baryons, so that the ratio of gas over total matter is 
$\omega_b=\Omega_b/\Omega_m$, and that the hydrogen abundance by mass has its primordial value, $X$. Under these assumptions

\begin{equation}
n_{\rm H}(r)=n_{{\rm H},0}\left(\frac{R_h}{r}\right)^\alpha;\quad 
n_{{\rm H},0}=X\,\omega_b\,\frac{\rho_h}{m_{\rm H}};\quad
\rho_h\approx \frac{200}{3}\rho_c\,.
\label{eq:profile}
\end{equation}
Here, $m_{\rm H}$ is the mass of a hydrogen atom and $n_{{\rm H},0}$ is the hydrogen number density at the virial radius\footnote{The factor 200/3 that relates $\rho_h$ to the value of the critical density, $\rho_c$, assumes that $\alpha=2$ rather than our preferred value of 2.2. The steeper profile leads to an apparent divergence of the enclosed mass at the centre. However, in reality, the power-law profile ceases to hold inside the accretion shock. Such a normalization should be approximately consistent with the similarity solution.}; in the figures we use \cite{Planck14}'s values of the cosmological parameters. In practise, we add the mean density to $n_{\rm H}$, to avoid the halo's density falling below the mean\footnote{This has a minor effect which can just be noticed outside the virial radius of the $M_h=10^9{\rm M}_\odot$ halo in Fig.~\ref{fig:radial}.}.

The cosmological growth of a dark matter halo is described by
\begin{equation}
M_h(z) = M_{h,0}\,m_h(z);\quad 
m_h(z)=(1+z)^a\,\exp(-bz)\,,
\label{eq:correaM}
\end{equation}
where $M_h$ is the virial mass of the halo at redshift $z$, $M_{h,0}$ is the mass of the halo at $z=0$, and $a\approx 0.24$ and $b\approx 0.75$ are fits by \cite{Correa15a, Correa15b} to simulations.

Given the total hydrogen profile, we proceed to compute the {\em neutral} hydrogen density profile, $n_{\rm HI}(r)$. The ionizing radiation from the UV-background will photo-ionize the intergalactic medium ({\sc igm}) as well as gas in the outskirts of the halo. As the gas gets denser, the optical depth to the ionizing radiation increases, and eventually hydrogen starts to self-shield. To describe this, we assume that (1) hydrogen is in photo-ionization equilibrium at a constant temperature of $T=10^4{\rm K}$, (2) collisional ionization can be neglected, (3) the presence of helium and of all other elements can be neglected, (4) the frequency dependence of the photo-ionization cross-section can be neglected, and (5) the optical depth
can be approximated by its value in a slab (rather than spherical) geometry\footnote{See \citealt{Zheng02, Sykes19} for calculations in spherical geometry.}. Under these assumptions, the neutral hydrogen fraction, $x$, at radius $r$, is given by 
\begin{eqnarray}
x(r) &\equiv &\frac{n_{\rm HI}(r)}{n_{\rm H}(r)}=\frac{\alpha_B}{\Gamma(r)}n_{\rm H}(r)\nonumber\\
 \Gamma(r)&=&\Gamma_0\exp(-\tau(r))\nonumber\\
 \tau(r)&=&\sigma_{\nu_{\rm th}}\int_r^\infty x(r')\,n_{\rm H}(r')\,dr'\,,
\label{eq:xneut}
\end{eqnarray}
where\footnote{We take $\alpha_B~=~1.269\times~10^{-13}~\eta^{1.503} / (1 + (\eta/0.522)^{0.407})^{1.923}~{\rm cm}^3{\rm s}^{-1}$ where $\eta=315614{\rm K} / T$ with $T=10^4{\rm K}$ and $\sigma_{\nu_{\rm th}}=6.63\times 10^{-18}{\rm cm}^2$.} $\alpha_B$ is the \lq case-B\rq\ recombination coefficient evaluated at a temperature of $T=10^4{\rm K}$, $\sigma_{\nu_{\rm th}}$ the photo-ionization cross-section at the ionization threshold, and $\Gamma_0$ the value of the photo-ionization rate in the {\sc igm}, for which we will take the values as a function of redshift computed by \cite{Haardt12}. Rather than performing the integral for the optical depth to infinity, we integrate inwards from $r=10\times R_h$, setting $\Gamma=\Gamma_0$ at $r=10\times R_{\rm h}$. Since the gas is highly ionised outside $R_h$, the exact value taken for this integration limit is mostly irrelevant, provided it is larger than $R_h$. Obtaining the run of $x$ versus $r$ numerically is straightforward, it is also  easy to understand the result: in the outskirts of the halo the gas is highly ionized so that $\tau\approx 0$. Once $\tau\approx 1$, the reduction in ionizing flux $\propto\exp(-\tau)$ increases very rapidly with decreasing $r$, hence the neutral fraction increases rapidly from very small to $x\approx 1$ over the narrow extent of the ionization front. Inside the front, $x\approx 1$. Therefore we expect that

\begin{equation}
\begin{cases}
x(r)\approx ({\alpha_B}/{\Gamma_0})n_{\rm H}(r)\rightarrow n_{\rm HI}(r)\approx ({\alpha_B}/{\Gamma_0})\,n_{\rm H}(r)^2\,\,\, \hbox{for $r>r_I$}\\
x(r)\approx 1\rightarrow n_{\rm HI}(r)\approx n_{\rm H}(r)\,,\quad \hbox{for $r<r_I$}\,,
\end{cases}
\label{eq:xprofile}
\end{equation}
where $r_I$ is the location of the ionization where $\tau\approx 1$. The value of $r_I$ can be estimated in the slab geometry by using the first expression for the neutral hydrogen density, and
solving $\sigma_{\rm th}\int_{r_I}^{10R_h}n_{\rm HI}(r')\,dr'=\tau_{r_I}=1$ for $r_I$, as we discuss in more detail below.

The run of total density and neutral hydrogen density obtained by numerically integrating Eq.~(\ref{eq:profile}) together with Eq.~(\ref{eq:xneut}) for halos with a range of masses is shown at redshift $z=3$ in panel (a) of Fig.\ref{fig:radial}, where the {\em solid lines} show $n_{\rm H}(r)$ and the {\em dotted lines} $n_{\rm HI}(r)$; different colours correspond to different halo masses. As expected, $n_{\rm HI}\approx n_{\rm H}$ close to the centre, and $n_{\rm HI}\propto n_{\rm H}^2$ in the outskirts: the inner most red-dashed line is $\propto n_{\rm H}$ and the outer most red-dashed line is $\propto n_{\rm H}^2$ to guide the eye. The location of the ionization front occurs where $\tau\approx 1$ as expected; the radius $r_I$ of the front is between 0.1 and 0.25 times the virial radius for the halos shown ($r_I/R_h$ is smaller for lower mass halos). 

The column-density through the neutral gas is
\begin{eqnarray}
N_{\rm HI}(b)&=&\int_{-\infty}^\infty n_{\rm HI}(r=(b^2+z^2)^{1/2})\,dz\nonumber\\
&=&2\,\int_b^\infty n_{\rm HI}(r) \frac{r}{(r^2-b^2)^{1/2}}\,dr\,,
\label{eq:cdens}
\end{eqnarray}
where $b$ is the impact parameter; in practise we use $10\,R_h$ for the upper integration limit rather than infinity. Since $n_{\rm HI}(r)$ is to a good approximation a power-law,
$n_{\rm HI}(r)\approx r^{-\beta}$ (where $\beta=\alpha$ for $r<r_I$ and $\beta=2\alpha$ for $r>r_I$), we expect that approximately
\begin{equation}
N_{\rm HI}(b)\approx 2\,n_0\,b^{1-\beta}\,R_h^{\beta}\int_0^\infty \frac{1}{(1+y^2)^{\beta/2}}\,dy\,\approx b^{1-\beta}\,R_h^{\beta}\,,
\label{eq:cdens_approx}
\end{equation}
that is, also a power-law in $b$, with the dependence on halo mass a power-law as well, since $R_h\propto M_h^{1/3}$. Here, $n_0=x\,n_{\rm H}$ with $x$ the neutral fraction at the virial radius for $r>r_I$, and $x=1$ for $r<r_I$, as in Eq.~(\ref{eq:xprofile}). $N_{\rm HI}(b)$, integrated numerically, is shown in panel (b) of Fig.~\ref{fig:radial} for the halos shown in panel (a); the power-law relation captures the shape of the $N_{\rm HI}$ profile well, as expected. Notice also that $N_{\rm HI}$ increases rapidly by a factor of several hundred at the location of the ionization front, $r=r_I$, where $x$ changes rapidly from $x\ll 1$ to $x\approx 1$.

We estimate the approximate location of the ionization front, $r_I$, by assuming that the gas outside $r_I$ is highly ionised as in Eq.~(\ref{eq:xprofile}). Integrating from $r_I$ outwards, the optical depth to infinity is
\begin{eqnarray}
\tau_I 
&=& \sigma_{\nu_{\rm th}}\int_{r_I}^\infty n_{\rm HI}\,dr
=\frac{\sigma_{\nu_{\rm th}}\,\alpha_B}{\Gamma_0}\int_{r_I}^\infty n_{\rm H}(r)^2\,dr\nonumber\\
&=&\frac{\sigma_{\nu_{\rm th}}\,\alpha_B\,n_{{\rm H},0}^2}{\Gamma_0\,(2\alpha-1)}\left(\frac{R_h}{r_I}\right)^{2\alpha-1}\,R_h\sim 1\,,
\end{eqnarray}
so that
\begin{equation}
r_I=\left(\frac{\sigma_{\nu_{\rm th}}\alpha_B\,n_{{\rm H},0}^2\,R_h}{(2\alpha-1)\Gamma_0}\right)^{1/(2\alpha-1)}\,R_h\,.
\label{eq:rI}
\end{equation}
The locations corresponding to this value of $r_I$ are indicated in Fig.~\ref{fig:radial} using {\em filled diamonds}. Defined in this way, the column-density measured outwards from $r_I$ to infinity is $N_{\rm HI}=\tau/\sigma_{\rm th}$ with $\tau=1$, corresponding to $\log N_{\rm HI}[{\rm cm}^{-2}]\approx 17.2$ - the characteristic column-density of Lyman-limit systems. However, the column-density of the halo for $b=r_I$ is of course a bit higher because of the different geometry of the two sight lines: radially outwards for the definition of $r_I$ versus hitting the absorber at a given impact parameter for an absorber.

We can use the $N_{\rm HI}-b$ relation to compute the cross-section, $\sigma(N_{\rm HI})$,
within which the column density is larger than some value,  $\sigma=\pi b^2\propto N_{\rm HI}^{2/(1-\beta)}$. Its derivative, $d\sigma/dN_{\rm HI}$, is closely related to the shape of the \cddf, $f(N_{\rm HI})$. To see this, 
start from the average number of times, $dN$, that a sight-line with proper length length $dl$
intersects absorbers with proper cross-section $\sigma$ and co-moving number-density $n(z)$,
\begin{eqnarray}
dN &=& n(z)\,(1+z)^3\,\sigma\,dl=n(z)\,(1+z)^3\,\sigma\,\frac{c\,dz}{(1+z)H(z)}\nonumber\\
&\equiv& \frac{c}{H_0}\,n(z)\,\sigma\,dX\nonumber\\
dX&\equiv &\frac{H_0(1+z)^2}{H(z)}\,dz\,,
\label{eq:dN}
\end{eqnarray} 
where $H(z)$ is the Hubble constant at redshift $z$, $H_0=H(z=0)$ and the dimensionless quantity $dX$ is called the \lq absorption distance\rq; the factor $(1+z)^3$ converts the co-moving density of absorbers, $n(z)$, to a proper number density (\citealt{Bahcall69} or see for example \cite{Fumagalli11}, \S 5.1). Therefore the number of times a sight-line intersects an absorber with a given column-density, due to a halo with a given mass per dex in halo mass, is given by
\begin{equation}
g(M_h, N_{\rm HI})\equiv \frac{d^3N}{dN_{\rm HI}d\log M_hdX} = \frac{c}{H_0}\frac{dn}{d\log M_h}\,\frac{d\sigma(M_h, N_{\rm HI})}	{dN_{\rm HI}}\,.
\label{eq:gCDDF}
\end{equation}

In this paper we use the \col\ {\sc python} package described by \cite{Diemer18} to compute
the number density of halos per dex in halo mass (the halo mass function)
$dn/d\log M_h$, stipulating a cosmology with the cosmological parameters from \cite{Planck14}. We selected the \cite{Reed07} fitting function to the mass-function as implemented in \collosus.

The \cddf\ - the number of absorbers with a given column-density per absorption distance $dX$ - is the integral of the function $g$ over halo mass,
\begin{equation}
f(N_{\rm HI}) \equiv \frac{d^2N}{dN_{\rm HI}dX}=\int_{-\infty}^{\infty} g(M_h, N_{\rm HI})\,d\log M_h\,.
\label{eq:fN}
\end{equation}

The contribution of halos of given mass to the \cddf, $g(M_h, N_{\rm HI})$,  is plotted in panel (c) of Fig.~\ref{fig:radial}. Its dependence on column-density is due to the factor $d\sigma(M_h, N_{\rm HI})/dN_{\rm HI} \propto N_{\rm HI}^{-2/(\beta-1)-1}$, with the power-law approximation shown by {\em red-dashed} lines\footnote{As before, $\beta$ is the slope of the 
the radial neutral hydrogen density profile, $n_{\rm HI}\propto r^{-\beta}$, which is $\beta=\alpha$ when the gas is neutral and $\beta=2\alpha$ when it is highly ionized.}. The power-law shape of $g$ is therefore a consequence of the power-law profile of the density in the halos, shown in panel (a), with the change in slope a consequence of the transition from fully neutral in the centre of the halos, so that $n_{\rm HI}(r)\propto n_{\rm H}(r)$, to the highly ionized regime in the outskirts, where $n_{\rm HI}(r)\propto n^2_{\rm H}(r)$. The inflection around $N_{\rm HI}\sim 10^{20}{\rm cm}^{-2}$ occurs when the value of the impact parameter is (approximately) equal to $r_I$.

It is striking how at higher values of $N_{\rm HI}$, $N_{\rm HI}\sim 10^{21}{\rm cm}^{-2}$ say, the contribution of halos is almost independent of halo mass over 2 orders of magnitude in $M_h$. The underlying reason for this is as follows: firstly, we will neglect the contribution of the ionised halo of gas to the column-density - this is an excellent approximation at sufficiently high $N_{\rm HI}$. Therefore, at impact parameter $b\ll r_I$, we find that
\begin{eqnarray}
N_{\rm HI}(M_h,b,z)&\approx &2\,n_{{\rm H},0}(z)\,R^{\alpha}_h(M_h,z)\,b^{1-\alpha}\,I\nonumber\\
 I&\equiv &\int_0^{\left((r_I(z)/b)^2-1\right)^{1/2}}\,\frac{dx}{(1+x^2)^{\alpha/2}}\,,
\label{eq:NHI}
\end{eqnarray}
where we have explicitly written-out the redshift dependence for later use. We can solve this equation for the value of the impact parameter $b$ at which a sight-line through a halo of mass $M_h$ has column-density $N_{\rm HI}$ or greater, and calculate the corresponding cross section 
$\sigma\equiv \pi b^2$, 
\begin{eqnarray}
\sigma(N_{\rm HI}, M_h, z)&=&\pi\,\left(2\,n_{{\rm H},0}(z)\right)^{2/(\alpha-1)}\,R_h^{2\alpha/(\alpha-1)}\nonumber\\
&\times & N_{\rm HI}^{-2/(\alpha-1)\,}  I^{2/(\alpha-1)}\,.
\label{eq:sigma1}
\end{eqnarray}
To gain insight and obtain a simpler analytical scaling relation, we further neglect the somewhat awkward dependence of $\sigma$ on the integral $I$, in which case $d\sigma/dN_{\rm HI}\propto R_h^{2\alpha/(\alpha-1)}\propto M_h^{2\alpha/3(\alpha-1)}$, where we have used the fact that $n_{{\rm H}, 0}$ depends on redshift but not on $M_h$. Finally, we find that the contribution of halos of mass $M_h$ to the \cddf\ depends on halo mass as
\begin{eqnarray}
g(M_h, N_{\rm HI}\sim 10^{21}{\rm cm}^{-2})&\propto &\frac{dn}{d\log M_h}\,\frac{d\sigma}{dN_{\rm HI}}\nonumber\\
&\approx & M_h^{-1.1}\,M_h^{2\alpha/3(\alpha-1)}\nonumber\\
&\approx &M_h^{0.13}\,,
\label{eq:gz}
\end{eqnarray}
where we have taken a value of $-1.1$ for the slope of the halo mass function at low masses at $z=3$, and our default value of $\alpha=2.2$ for the slope of the density distribution within a given halo. The contribution of halos to the \cddf\ is thus a slowly increasing function of $M_h$ at high $N_{\rm HI}$, increasing by a factor of $\sim 2$ for $M_h$ increasing by a factor of 100 from $M_h=10^{10}{\rm M}_\odot$ to $10^{12}{\rm M}_\odot$. At even higher mass, the contribution eventually drops due to the exponential factor in the halo mass function. The scaling exhibited by Eq.~(\ref{eq:gz}) explains the weak dependence of the \cddf\ on halo mass.
We will verify the scaling in more detail numerically below, including the impact of the integral $I$.

We could now integrate Eq.~(\ref{eq:gCDDF}) over halo mass to obtain the \cddf, as in Eq.~(\ref{eq:fN}). Before doing so, we should be more careful in considering for which halos our model assumptions hold.  The ionizing background will photo-evaporate gas from halos below some critical mass, $M_{\rm crit}(z)$, and such halos will not contribute to the \cddf.  Here, we will use the values of $M_{\rm crit}(z)$ obtained using numerical simulations by \cite{Okamoto08}. At {\em high} halo masses, $M_h\gg 10^{12}{\rm M}_\odot$, on the other hand, a lot of the halo gas is presumably shock heated rather than cold, in which case we should not neglect collisional ionization and additionally the radial density profile is likely to be affected. Fortunately above $z\sim 2$, such halos are relatively rare and their contribution to the integral in Eq.~(\ref{eq:fN}) is already exponentially suppressed due to the shape of the halo mass function, meaning we may not need to further suppress their contribution. We will therefore compute the \cddf\ as
\begin{equation}
f(N_{\rm HI}) \equiv \frac{d^2N}{dN_{\rm HI}dX}=\int_{\log M_{\rm crit}(z)}^{\infty} g(M_h, N_{\rm HI})\,d\log M_h\,,
\label{eq:fN2}
\end{equation}
where in practice we integrate up to halos of mass $10^{13}{\rm M}_\odot$. There remains one further subtlety regarding the definition of $M_h$. To compute the number density of halos, $dn/d\log M_h$, we use the \col\ routines of \cite{Diemer18}, which are fits to dark matter only ({\sc dmo}) simulations.  In such simulations, the mass of a halo, $M_h$, is the sum of the dark matter plus baryon mass. Therefore when comparing to hydrodynamical simulations, we should not compare the dark matter halo mass to $M_h$, but rather $\Omega_m/\Omega_{\rm dm}$ times the dark matter mass  (with, of course, $\Omega_m$ and $\Omega_{\rm dm}$ the mean cosmological mass and dark matter density in units of the critical density). Of course in reality, outflows of baryons due to feedback from star formation slow the growth in mass of a halo \cite[e.g.][]{Sawala15}, a process we will neglect in this analytical investigation. We are now in a position to compute the \cddf\ and study its evolution.

\subsection{The \cddf\ and its redshift evolution}
\label{sect:redshift}
\begin{figure*}
	\includegraphics[width=.95\textwidth]{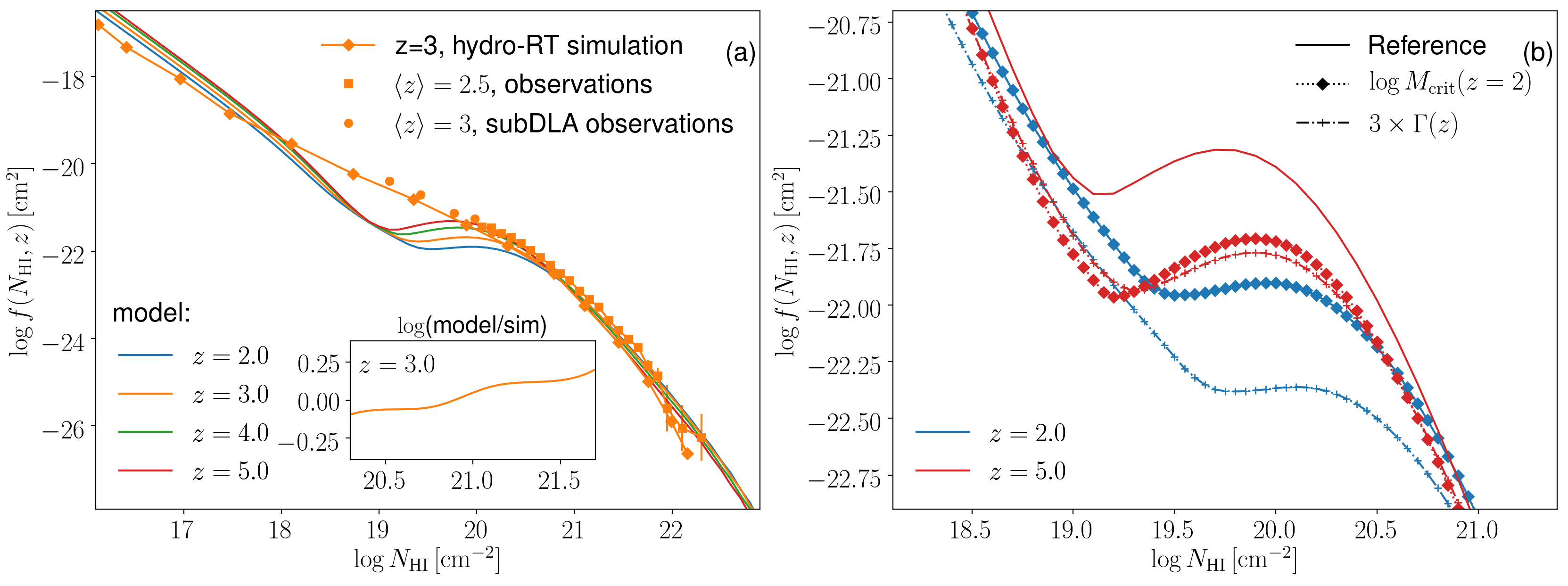}\hfill
	\caption{{\bf Panel (a)}: The model \cddf\  function $f(N_{\rm HI}, z)$, at various redshifts plotted in {\em solid lines}. Over plotted is $f( z=3)$ for the {\sc owl} simulations \protect\citep{Schaye10} presented by \protect\cite{Altay11} ({\em yellow line connecting diamonds}) and $f(\langle z\rangle=2.5)$ for \dla s measured by \protect\cite{Noterdaeme12} ({\em yellow squares}) and sub-DLAs from \protect\cite{Zafar13} ({\em yellow circles}), which both fall approximately on top of the yellow simulations line. The {\em inset} shows the logarithm of the ratio of model over simulation at $z=3$. {\bf Panel (b)}:  $f(N_{\rm HI}, z)$ at $z=2$ and $z=5$ for the reference model ({\em solid lines}). A variation on the reference model in which the critical mass, $M_{\rm crit}$ (below which halo gas is photo-evaporated and no longer contributes to $f$) does not evolve, $M_{\rm crit}(z)=M_{\rm crit}(z=2)$, is shown as {\em dotted lines connecting diamonds} (this variation overlaps with the reference model at $z=2$). Another variation in which the amplitude of the ionising background, $\Gamma$, is multiplied by a factor of 3 at all $z$ is shown as {\em dot-dashed lines connecting crosses}.
	}
	\label{fig:cddf}
\end{figure*}
\begin{figure*}	
	\includegraphics[width=.95\textwidth]{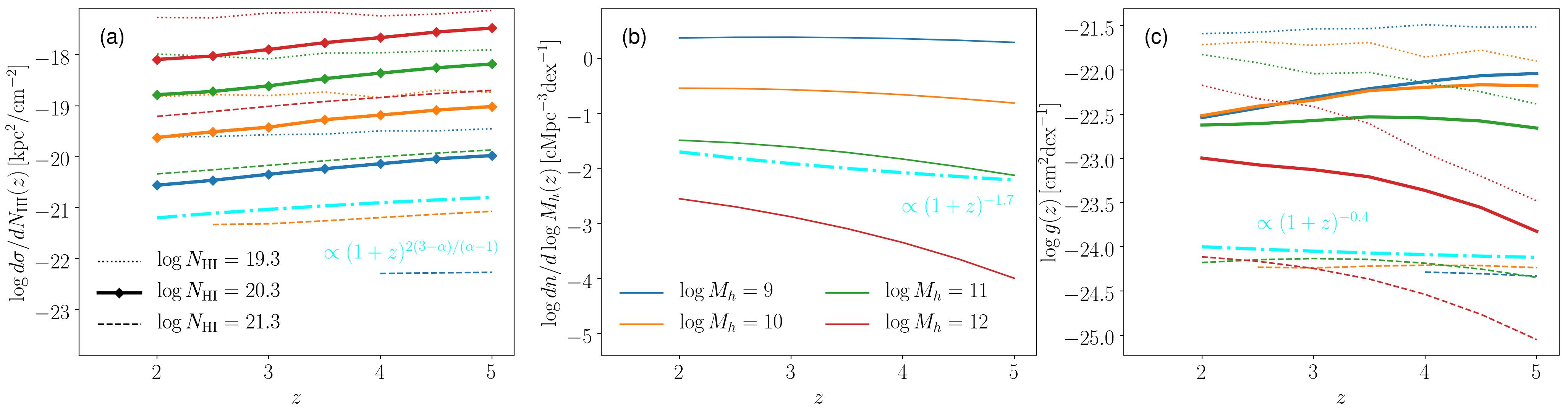}\hfill
	\caption{Dependence of the contribution of halos with given mass, 
		($\log M_h/M_\odot$=9, 10, 11 and 12 shown in red, green, yellow and blue, respectively)
		at a given column-density ($\log N_{\rm HI}/{\rm cm}^{-2}=19.3$, 20.3 and 21.3 as dotted, thick solid, and dashed, respectively) to the \cddf\ as a function of redshift, from Eq.~(\ref{eq:gCDDF}). {\bf Panel (a)}: cross-section, $d\sigma/dN_{\rm HI}$,
		as a function of $z$. {\bf Panel (b):} dark matter halo number density, $dn/d\log M_h$, as a function of redshift. {\bf Panel (c):} net contribution $g(N_{\rm HI}, M_h, z)$ of halos
		of mass $M_h$ to the \cddf\ as column density $N_{\rm HI}$, as a function of redshift. {\em Cyan dash-dotted lines} in each panel indicates the redshift scaling as annotated.
		Panels (a) and (b) show that the evolution of cross section and number density is relatively weak, as expected (see text). The trends are also in the opposite direction, leading to even weaker evolution of $g$, as shown in panel (c).}
	\label{fig:cddf_evolution}
\end{figure*}
\begin{figure}
	\includegraphics[width=.95\columnwidth]{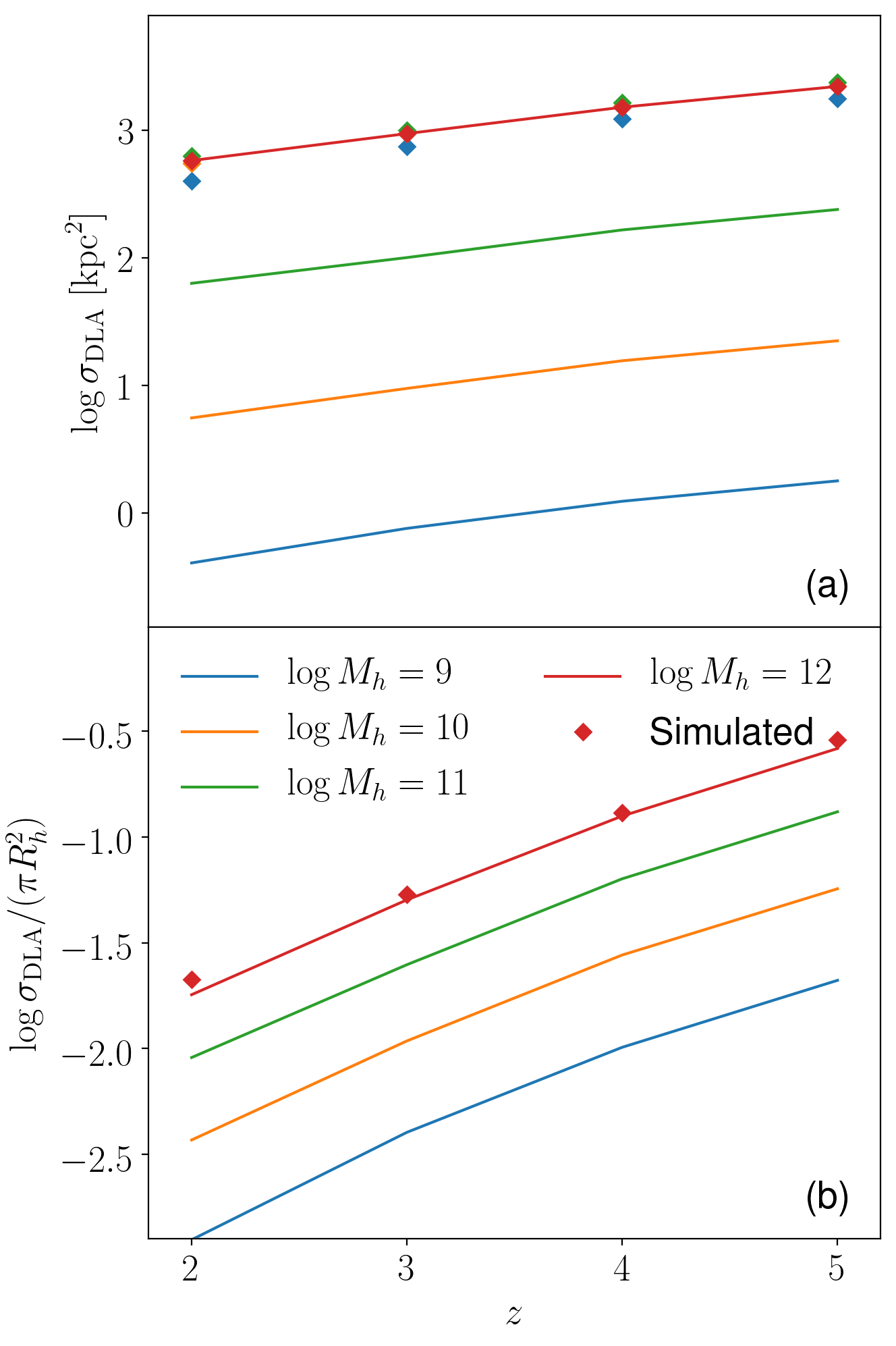}\hfill
	\caption{{\b Panel (a)}: {\em Coloured curves} are the model's \dla\ cross section, $\sigma_{\rm DLA}$, in proper units, as a function of redshift. Curves are coloured according to halo mass, as per the  legend in panel (b). {\em Filled diamonds} show $\sigma_{\rm DLA}$ scaled by $10^{12}{\rm M}_\odot/M_h$.
		{\bf Panel (b)}: \dla\ covering fraction, $\sigma_{\rm DLA}/(\pi R_h^2)$, as a function of redshift for
		halos of different mass (colours). The {\em filled red diamonds} are the covering factors computed using the fitting formula presented by \protect\cite{Rahmati15} from the \eagle\ simulations, for halos with mass $\ge 10^{12}{\rm M}_\odot$.}
	\label{fig:covering}
\end{figure}
\begin{figure*}	
	\includegraphics[width=.95\textwidth]{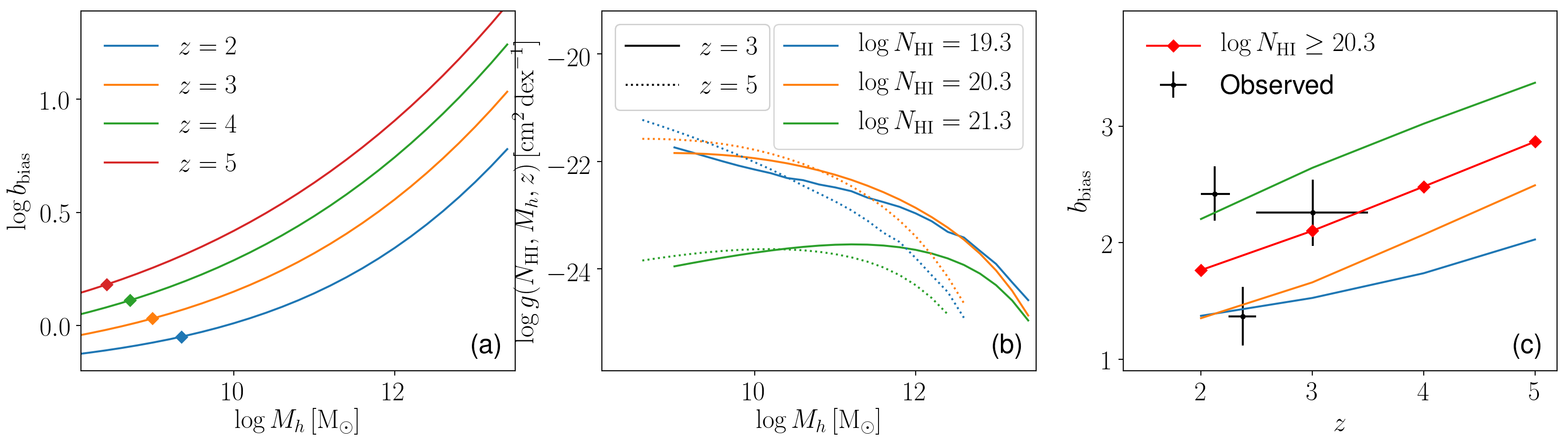}\hfill
	\caption{Evolution of \dla\ bias. {\bf Panel (a):} halo bias as a function of mass at different
		redshifts  (various colours), computed for the \protect\cite{Planck14} cosmology using the \col\ code of \protect\cite{Diemer18} with the \protect\cite{Tinker10} prescription. The {\em solid diamonds} indicate the critical mass, $M_{\rm crit}$, below which gas in halos is photo-evaporated according to the simulations of \protect\cite{Okamoto08}. Halos of lower mass are assumed not to contribute to the \cddf. {\bf Panel (b):} Differential contribution of halos of a given mass to the \cddf\ from Eq.~(\ref{eq:gz}) for three values of $N_{\rm HI}$ (different colours) and two redshifts (different line styles) as per the legend. Notice in particular that $g$ is depends weakly on halo mass over a large range in $M_h$. {\bf Panel (c):} bias of absorbers at a given value of the column-density, as per the legend in panel (b); the {\em red line} corresponds to \dla s. The {\em black squares} are the observed \dla\ bias measurements from \protect\cite{DLAbias18}.}
	\label{fig:bias}
\end{figure*}

Numerically integrating Eq.~(\ref{eq:fN2}) yields the \cddf\ with the result plotted in panel (a) of Fig.~\ref{fig:cddf} at four redshifts. The model (coloured curves for different redshifts) is compared to the results from a hydrodynamical simulation at $z=3$ ({\em yellow line connecting diamonds}) from the {\sc owls} project \citep{Schaye10}, post-processed with radiative transfer using the {\sc urchin} reverse ray-tracing code described by \cite{Altay13b}, as presented by \cite{Altay11}. Also shown as a {\em yellow diamonds with error bars} is the observed \cddf\ of \dla s at a mean redshift of $\langle z\rangle\approx 2.5$ as measured by \cite{Noterdaeme12} from the {\sc sdss} {\sc dr9} \citep{DR9}), and the \lq sub-DLAs\rq\ measured by \cite{Zafar13} at $z\sim 3$ ({\em yellow circles}). Simulation and model are in excellent agreement with the data (better than 50\%) above the \dla\ threshold, $N_{\rm HI}>10^{20.3}{\rm cm}^{-2}$. The model also agrees with the simulation at lower columns although not as well, with in particular the transition from mostly neutral absorbers ({\em i.e.} \dla s) to absorbers that are highly ionized in the model (sub-\dla s, $N_{\rm HI}\sim 10^{19-20.3}{\rm cm}^{-2}$ and Lyman-limit systems, $N_{\rm HI}\geq 10^{17.2}{\rm cm}^{-2}$) more  abrupt than in the simulation or in the data. However the overall agreement at higher column densities between the simple model and the simulations, as plotted in the inset of Fig.~\ref{fig:cddf}a,  is  encouraging. We note that the latter include molecule formation but the model does not (but see \S \ref{sect:molecules}).

What determines the shape of the \lq knee\rq\ feature at $N_{\rm HI}\sim 10^{20}{\rm cm}^{-2}$? As illustrated in Fig.~\ref{fig:radial}, when the impact parameter $b$ becomes smaller than $r_I$, the radius of the ionization front, the column density rapidly increases by $\sim 2$ orders of magnitude for a small change in $b$. For a {\em given} halo, this results in a nearly flat region in $g(N_{\rm HI}, M_h)\propto d\sigma/d N_{\rm HI}$, as seem in panel (c) of Fig.~\ref{fig:radial}: this is the origin of the knee. \cite{Zheng02} previously pointed-out the connection between the rapid onset of self-shielding in a spherical cloud and the appearance of a knee in the \cddf, see also \cite{Petitjean92}. With the \cddf, $f(N_{\rm HI})$, an integral over $g(N_{\rm HI}, M_h)$, the knee in $g$ results in a knee in $f$, however the shapes are different because the location of the knee is weakly dependent on $M_h$. We will return to this after first investigating evolution.

The observed \cddf\ evolves very weakly even over the extreme redshift range $z=0\rightarrow 5$.  Such weak evolution is reproduced by simulations, see {\em e.g} Fig.~1 in \cite{Rahmati13}, although the underlying reason for the near absence of evolution from $z=2\rightarrow 5$ has not been discussed in great detail. The model also shows very weak evolution, see panel (a) of Fig.~\ref{fig:cddf}. Given the weak dependence of $g$ on halo mass in the current model, weak evolution in $f$ requires weak evolution in $g$.

We therefore examine the predicted evolution of the function $g(N_{\rm HI}, M_h)\propto (d\sigma/dN_{\rm HI})\,(dn/d\log M_h)$ of Eq.~(\ref{eq:gz}) in more detail, starting from Eq.~(\ref{eq:sigma1}) for the evolution of the cross section, $d\sigma/d N_{\rm HI}$. Using $\sigma=\pi b^2$, we solve Eq.~(\ref{eq:sigma1}) for the value of the impact parameter $b$ for a sightline through a halo of mass $M_h$ to have a given column density $N_{\rm HI}$. Keeping $M_h$ and $N_{\rm HI}$ both constant, and neglecting the contribution from the integral $I$, the redshift dependence is due to the evolution of $n_{{\rm H}, 0}$ - the (total) hydrogen density at the edge of the halo, and $R_h$ - the virial radius of a halo of given mass $M_h$. This yields for the net redshift dependence of $\sigma$ and its derivative with respect to $N_{\rm HI}$
\begin{eqnarray}
\frac{d\sigma}{dN_{\rm HI}}
&\propto& n_{{\rm H},0}^{2/(\alpha-1)}\,R_h^{2\alpha/(\alpha-1)}\nonumber\\
&\propto&  (1+z)^{6/(\alpha-1)}\,(1+z)^{-2\alpha/(\alpha-1)}\nonumber\\
&\propto &(1+z)^{2(3-\alpha)/(\alpha-1)}\,.
\end{eqnarray}
This scaling is a result of halos becoming larger at later times, $R_h\propto (1+z)^{-1}$, but less dense, $n_{{\rm H}, 0}\propto (1+z)^{3}$. The density dependence is stronger and the cross section (at a given column density and given mass) {\em decreases} with time. 

However the (co-moving) number density of halos with mass $M_h$, $dn/d\log M_h$,  {\em increases} with time. In the Press-Schechter approximation, the increase is proportional to the linear growth rate, $D(z)$, which to a good approximation is
$D(z)\approx 1/(1+z)$ in the redshift range $z=2\rightarrow 5$. However, in the mass-range of interest, $9\le \log M_h/{\rm M}_\odot\le 13$, the evolution is slightly stronger, more like $dn/d\log M_h\propto (1+z)^{-1.7}$. 

The opposite evolution of the cross section and the number density of halos yields for the net evolution for the contribution of halos with given $M_h$ to the \cddf\ at a given (self-shielded) column,
\begin{eqnarray}
	g(N_{\rm HI}, z)&\propto &\frac{dn}{d\log M_h}\,\frac{d\sigma}{dN_{\rm HI}}\nonumber\\
	&\propto &(1+z)^{-1.7}\,(1+z)^{2(3-\alpha)/(\alpha-1)}\nonumber\\
	&\approx &(1+z)^{-0.4}\,.
	\label{eq:z-scaling}
\end{eqnarray}
Using the value at $z=3$ as a pivot point, $g(z=2)/g(z=3)\approx 1.1$, and $g(z=5)/g(z=3)\approx 0.86$.  Therefore as time progresses, more halos of a given mass $M_h$ appear at lower $z$ but the cross section of individual halos decreases. These two factors nearly compensate each other, resulting in weak evolution of the function $g$ - and explaining the weak evolution in the column-density distribution of \dla s, $f(N_{\rm HI})$.

These analytical estimates are tested in more detail in Fig.~\ref{fig:cddf_evolution}.
In panel (a) of that figure, we show the evolution of the cross section, $d\sigma/d N_{\rm HI}$
for halos with various masses (coloured lines), and for column densities of
$\log N_{\rm HI}[{\rm cm}^{-2}]$=19.3 (dotted), 20.3 (solid) and 21.3 (dashed lines). Evolution is relatively weak, following approximately the evolution $d\sigma/ dN_{\rm HI}\propto (1+z)^{2(3-\alpha)/(\alpha-1)}$ shown as the {\em dot-dashed cyan line}, for all $M_h$ and values of $N_{\rm HI}$ shown. In panel (b) we show the evolution of the co-moving number density, $dn/d\log M_h$, computed using \col\ \citep{Diemer18}. To guide the eye, we over plot as the {\em dot-dashed cyan line}, the scaling $\propto (1+z)^{-1.7}$, which captures at least approximately the evolution of the lower mass halos, underestimating the evolution for $\log M_h/{\rm M}_\odot=12$. Finally, in panel (c) we plot the evolution of the contribution of halos of given mass and given column density, $g(N_{\rm HI}, M_h)$. Given that the trends in cross section and number density are weak and in the opposite direction, the evolution in $g$ is even weaker, following reasonable well the trend $\propto (1+z)^{-0.4}$ predicted earlier for
self-shielded  column-density systems, and shown as the {\em dot-dashed cyan line}.

It is also worth exploring the evolution of the covering factor, the ratio $\sigma/(\pi R_h^2)$,
which follows from Eq.~(\ref{eq:sigma1}) and is plotted in Fig.~\ref{fig:covering}. 
The \dla\ cross section (in proper kpc$^2$) increases slowly with redshift at constant halo mass $M_h$, and is approximately $\propto M_h$, as can be seen in the {\em upper panel}. 
Because the virial radius decreases with $z$, the covering factor increases faster with $z$ than the cross section ({\em lower panel}). The values of the covering factor and their evolution are
in very good agreement with those computed by \cite{Rahmati15} for galaxies in the \eagle\ simulation \citep{Schaye15} with $M_h\ge 10^{12}{\rm M}_\odot$ ({\em filled red diamonds}).
Another comparison is with the high resolution simulations of \cite{Faucher11} which yield a covering factor of $\sigma_{\rm DLA}/(\pi R_h^2)$ of a $\sim 3-10$ percent for $M_h\sim 2-3\times 10^{11}{\rm M_\odot}$ at $z=2$ - values comparable to what we find here\footnote{
The simulations by \citealt{Faucher15} that include feedback are claimed to give a far higher covering factor of 0.2-0.4 for such halos.}. 

According to Eq.(\ref{eq:sigma1}), $\sigma\propto R_h^{2\alpha/(\alpha-1)}\propto M_h^{2\alpha/(3(\alpha-1))}\approx M_h^{1.22}$. We have multiplied the values of $\sigma$ for each halo by $(10^{12}{\rm M_\odot}/M_h)^{1.22}$ and over plotted them on Fig.\ref{fig:covering} as {\em filled diamonds}. The fact that these symbols fall nearly on top of the $M_h=10^{12}{\rm M}_\odot$ values demonstrates that this scaling works very well.
The value of the exponent $\beta\sim 1.22$ in $\sigma\propto M_h^{\beta}$ is a bit larger than found in the numerical simulations presented by \cite{Bird14} (who finds that $\beta<1$  depending on the feedback strength in their simulations) and more comparable to that inferred from the observed DLA bias by \cite{Font12, DLAbias18}, who prefer $\beta\sim 1.1$. With $\beta=1.22$, the \dla\ {\em covering factor} scales as $\sigma/R_h^2\propto M_h^{0.56}$ in our model. 

Two other trends are worth noting in panel (c) of Fig.~\ref{fig:cddf_evolution}. Firstly, at high column densities ($\log N_{\rm HI}\,[{\rm cm}^{-2}]=21.3$), halos of $\log M_h/{\rm M}_\odot=10-12$ contribute about equally to $g$ (see the dashed lines). As discussed before, the dependence of $g$ on $N_{\rm HI}$ is weak at high column, see Eq.~(\ref{eq:gz}). Secondly, the dependence becomes stronger at lower columns, with lower mass halos contributing significantly more particularly at higher $z$ (contrast the dotted and the solid lines).

The second observation brings us back to the shape of the knee in $f(N_{\rm HI})$. At higher $z$, lower mass halos contribute relatively more to $f(N_{\rm HI})$ per decade in halo mass. In addition, the value of $M_{\rm crit}$ - the halo mass below which gas is evaporated from halos by photo-heating by the ionising background - is lower at higher $z$, meaning lower mass halos contribute even more to $f$ at $z=5$ compared to $z=2$. The result is a noticeable increase in the number density of absorbers slightly below the \dla\ threshold moving the location of the knee to lower column densities: this increase is due to an increase in the contribution of lower mass halos.  The consequence of this is best appreciated by comparing $f(N_{\rm HI}, z)$ at $z=5$ (red lines) versus at $z=2$ (blue line) in panel (a) of Fig.~\ref{fig:cddf}: there are considerably more sub-\dla s (with $N_{\rm HI}$ slightly below $10^{20.3}{\rm cm}^{-2}$) at $z=5$ compared with $z=2$. Such a change in shape is also apparent in simulations (see e.g. Fig.~1 in \citet{Rahmati15}, see also Sykes et al. {\em in prep.}). The data also suggest an increased abundance of subDLAs with increasing $z$ \citep{Zafar13}.

To illustrate the impact of $M_{\rm crit}$ on the shape of the knee, we plot in panel (b) of Fig.~\ref{fig:cddf} the \cddf, $f(N_{\rm HI}, z)$, for $z=2$ and $z=5$, comparing the default model in which $M_{\rm crit}$ evolves (solid lines) with a variation in which the critical mass is kept fixed at its value at $z=2$ (dotted lines connecting diamonds). These models obviously fall on top of each other at $z=2$, but the shape of the knee for the $z=5$ curves changes significantly - resulting in a large decrease in the number density of sub-\dla s. The panel also shows another model variation, in which
$M_{\rm crit}$ evolves as before, but the amplitude $\Gamma$ of the ionising background is multiplied by a factor of 3 at all $z$ above its value in the default model (that of \cite{Haardt12}; 
this variation is shown as {\em dot-dashed line connecting crosses}). Increasing $\Gamma$ changes the shape of the knee in a similar way as increasing $M_{\rm crit}$,  but the two variations do change the shape of the \cddf\ differently at lower column-densities. If it were possible to constrain $\Gamma$ robustly by observations at lower columns, for example in the Lyman-$\alpha$ forest \cite[e.g.][]{Becker13}, then evolution of the shape of the knee could be used to test models for the evolution of $M_{\rm crit}$. 
In fact, it is likely easier to constrain robustly the {\em evolution} of $\Gamma$ rather than its amplitude at any $z$ - this might be enough to test whether $M_{\rm crit}$ evolves as expected. Other effects may also play a role in shaping the knee, for example the temperature of the absorbing gas \citep{McQuinn11} or the hardness of the ionizing radiation which will affect the width of the ionization front.  Finally, we note the relatively substantial effect of the value of $M_{\rm crit}$ on the number density of Lyman limit systems \cite[see also][]{Fumagalli13}.

Armed with an estimate of how much halos of given mass contribute to the \cddf, it is now straightforward to compute the bias of \dla s, to which we turn next.

\subsection{\dla\ bias}
\label{sect:bias}
\begin{figure*}
	\includegraphics[width=.95\textwidth]{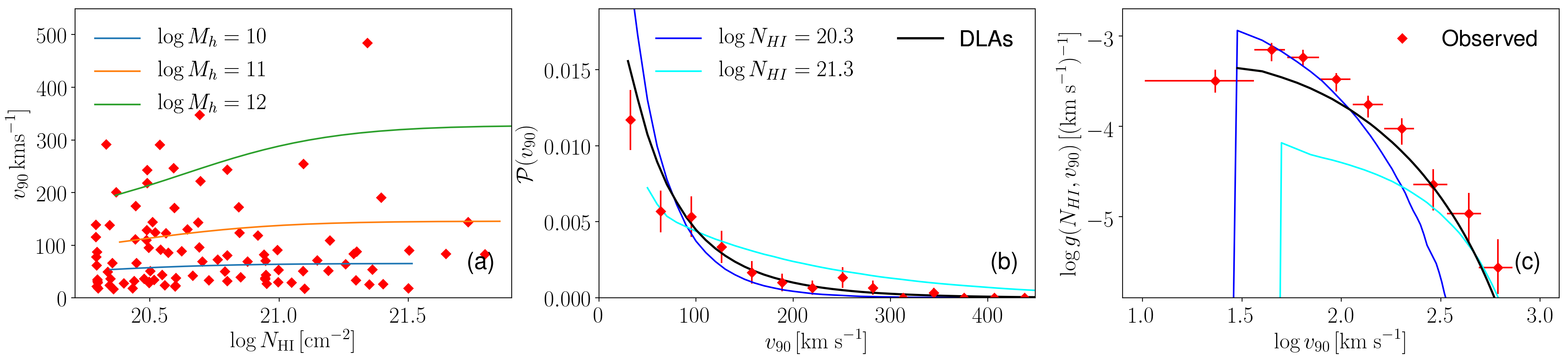}\hfill
	\caption{\dla\ line width, $v_{90}$, at redshift $z=3$. {\bf Panel (a)}: $v_{90}$ versus
		column density, $N_{\rm HI}$. {\em Lines} are the model prediction, with different colours referring to different halo masses, $M_h$ in solar masses, as indicated in the legend.  Along a curve for a given halo mass, impact parameter decreases with increasing $N_{\rm HI}$. {\em Red squares} are the observations by \protect\cite{Neeleman13}. {\bf Panel (b):} Fraction of \dla s with a given
		line width. {\em Lines} are the model prediction, with {\em blue} and {\em cyan lines} for column densities of $\log N_{\rm HI}[{\rm cm}^{-2}]=20.3$ and 21.3, respectively, and the {\em thick black line} the model prediction for \dla s, $\log N_{\rm HI}[{\rm cm}^{-2}]>20.3$. {\em Red symbols} correspond to the data of \protect\cite{Neeleman13} as in panel (a), plotted with Poisson error bars. Each histogram is normalised to unity. {\bf Panel (c):} Number density of absorbers with a given line width. Lines are the model prediction from Eq.~(\ref{eq:gv90}), with the same colouring as in panel (b). {\em Red symbols with error bars} are the data from \protect\cite{Prochaska03} as plotted in Fig.~9 of \protect\cite{Pontzen08}, for \dla s in the redshift interval $[1.6, 4.5]$. }
	\label{fig:v90}
\end{figure*}

The term {\em bias} was famously coined by \cite{Davis85} to describe the fact that observed galaxies are more strongly clustered than the mass seen in numerical simulations, with \cite{Kaiser84} showing that such biasing naturally follows from the assumption that galaxies form at local maxima in a Gaussian density field. Mathematically, the bias factor\footnote{We use the symbol $b_{\rm bias}$ to denote the bias factor to avoid confusion with the impact parameter, $b$.}  $b_{\rm bias}$ of a population of objects - for example halos with a given mass -  can be defined as the ratio of their power-spectrum, $P(M_h, z)$, to that of the mass, $P_m$: $b^2_{\rm bias}(M_h, z)\equiv P(M_h, z)/P_m(z)$. The \col\ tool of \cite{Diemer18} implements several methods for estimating $b_{\rm bias}$ for halos of a given mass at a given redshift. Below we use \cite{Diemer18}'s implementation of the model by \cite{Tinker10}. Given the function $g(N_{\rm HI}, M_h, z)$ and this bias function $b_{\rm bias}(M_h,z)$, we compute the bias of lines with a given value of the column density as  (see also \cite{Padmanabhan16})
\begin{equation}
b_{\rm bias}(N_{\rm HI}, z) = 
\frac{
\int_{\log M_{\rm crit}(z)}^{\infty}
d\log M_h\,\left\{b(M_h,z)\times g(N_{\rm HI}, M_h, z)\right\}
}
{
\int_{\log M_{\rm crit}(z)}^{\infty}
d\log M_h\,\left\{g(N_{\rm HI}, M_h, z)\right\}
}\,,
\end{equation}
and the bias of \dla s as
\begin{eqnarray}
&\hphantom{.}&\hspace{-1cm}b_{\rm bias}({\rm DLA}, z)=\nonumber\\
&\hphantom{..}&\hspace{-2cm}
\frac{
\int_{20.3}^{\infty} d\log N_{\rm HI}\int_{\log M_{\rm crit}(z)}^\infty d\log M_h\,
{\cal F}_1(M_h, N_{\rm HI}, z)
}
{ 
\int_{20.3}^{\infty} d\log N_{\rm HI}\int_{\log M_{\rm crit}(z)}^\infty d\log M_h\,
{\cal F}_2(M_h, N_{\rm HI}, z)
}\,.\nonumber\\
{\cal F}_1(M_h, N_{\rm HI}, z) &=& b(M_h, z)\times N_{\rm HI}\times g(N_{\rm HI}, M_h, z)\nonumber\\
{\cal F}_2(M_h, N_{\rm HI}, z) &=& N_{\rm HI}\times g(N_{\rm HI}, M_h, z)\,.
\end{eqnarray}
The results are illustrated in Fig.~\ref{fig:bias}. In panel (a) we plot the bias function $b_{\rm bias}(M_h,z)$  computed using \col\ \citep{Diemer18}. Superposed as {\em filled diamonds} are the values of the critical mass, $M_{\rm crit}$, computed from \cite{Okamoto08} for reference. In panel (b) we plot the function $g(N_{\rm HI}, M_h, z)$ for three values of $N_{\rm HI}$ ({\em colours}). As we noted and explained before, the function $g$ depends relatively weakly on $M_h$ over a large range in halo mass. In our model, it falls abruptly to zero for halos with mass $M_h<M_{\rm crit}$, by construction, and at high mass decreases exponentially due to the exponential cut-off in the halo mass function. A comparison between the {\em solid} and {\em dotted lines}, which correspond to redshifts $z=3$ and $z=5$, respectively, shows how the function $g$ evolves. $M_{\rm crit}$ decreases with increasing $z$ - so that lower mass halos contribute more to $g$ at higher $z$. Additionally, the location of the exponential cut-off in the mass function also decreases with increasing $z$ and consequently the relative contribution of lower mass halos to the \cddf\ increases with $z$.

Combining panels (a) and (b) yields the bias of absorbers, plotted in panel (c). In addition to the bias of the absorbers shown in panels (a) and (b), we also plot the bias of \dla s ({\em red line connecting solid diamonds}). The bias increases with $z$ for all column-densities shown in the panel. So, even though there is a tendency for lower mass halos to increase their contribution to the cross section with increasing redshift (which would decrease $b_{\rm bias}$), the bias of such halos {\em increases} even more rapidly with $z$, leading to a net increase in $b_{\rm bias}$.
We also note that the bias increases with column density, which is not surprising, but at $z\sim 2$, the bias is approximately the same for $\log N_{\rm HI}\,[{\rm cm}^{-2}]=19.3$ and 20.3 (blue and orange lines, respectively).  Since bias increases with column density,
the bias of \dla s, $b_{\rm bias}({\rm DLA}) > b_{\rm bias}(\log N_{\rm HI}\,[{\rm cm}^{-2}]=20.3)$. For reference we over plot the bias measurements of \dla s from \cite{DLAbias18} as {\em black symbols with error bars}\footnote{These correspond to the red filled circles in Fig.~4 of that paper.}. Although the agreement is not perfect, our \dla\ model yields reasonably high  values of $b_{\rm bias}$, even though low-mass halos contribute significantly. 

\subsection{\dla\ line widths}
\label{sect:v90}
To measure the velocity extent of \dla s expected in the model, we consider sight lines that intersect neutral gas. In the current model, those have impact parameter $b\le r_I$, the radius of the ionization front. We further assume that the gas is flowing radially with hydrogen accretion rate 
\begin{equation}
\dot M_{\rm H} = \omega_b\,X\,\dot M_h\,,
\end{equation}	
where we use Eq.~(\ref{eq:profile}) for the profile and Eq.~(\ref{eq:correaM}) for the halo accretion
rate. Using the continuity equation then yields for the radial inflow velocity of the hydrogen gas
\begin{eqnarray}
v_r &=& \frac{\dot M_{\rm H}}{4\,\pi r^2\,m_{\rm H}\,n_{\rm H}(r)}\nonumber\\
&=&\frac{1+z}{10\,\Omega_m}\,\,\left(b-\frac{a}{1+z}\right)\,\left(\frac{r}{R_h}\right)^{\alpha-2}\,v_h\nonumber\\
&\approx& {210}{\rm km~s}^{-1}\,\left(\frac{M_h}{10^{12}{\rm M}_\odot}\right)^{1/3}\,\left(\frac{1+z}{4}\right)^{3/2}\,\left(\frac{r}{R_h}\right)^{\alpha-2}\,,\nonumber\\
\label{eq:vr}
\end{eqnarray}
where the numerical values use the \cite{Planck14} values of the cosmological parameters. This infall velocity depends little on radius in the cases we are interested, {\em i.e.} when the slope of the density profile $\alpha\approx 2$. The component of the velocity along the line-of-sight is $v_z=\cos(\theta)\,v_r$, where $\cos(\theta)=z/r$ depends on the distance $r$ to the centre and on $z=(r^2-b^2)^{1/2}$. Given that $v_r$ only depends weakly on $r$ means that absorption in neutral gas occurs over a velocity range of approximately $[-v_{\rm HI}, v_{\rm HI}]$, where $v_{\rm HI}=(1-b^2/r_I^2)^{1/2}\,v_r(r=r_I)$, where
$v_r(r=r_I)$ is the value of the infall velocity at the location $r_I$ of the ionization front.

Observationally, the line width associated with a \dla\ is quantified by a velocity called $v_{90}$, defined such that 90 per cent of the optical depth of a line associated with a metal transition occurring in neutral gas is enclosed by this velocity interval. To connect this to our velocity profile, we will identify $v_{90}$ with $2\,v_{\rm HI}$,
\begin{equation}
v_{90} = 2 v_r(r_I)\,(1-\frac{b^2}{r_I^2})^{1/2}\,,
\label{eq:v90init}
\end{equation}
which is thus the velocity extent\footnote{With the factor 2 accounting for the gas falling away from as well as well as towards the observer.} of neutral gas at impact parameter $b$ for a halo of given mass, $M_h$. As we will show below, there is a (relatively weak) dependence of $v_{90}$ on column-density for any given halo, since at low columns the sight line grazes the ionization front at $r=r_{I}$, gas flows mostly perpendicular to the sight line, yielding a low value of $v_{90}$. At low $b$, the column-density is near maximum, gas flows along the sight line, and hence $v_{\rm HI}$ is maximal. This geometric dependence does not translate directly into a correlation between $N_{\rm HI}$ and $v_{\rm 90}$ {\em for a population} of \dla s, as we show below.

We obtain the joint probability distribution function ({\sc pdf}) for lines to have a given value of column-density and $v_{90}$ starting from Eq.~(\ref{eq:v90init}), by
changing variables from $b$ to $N_{\rm HI}$ using Eq.~(\ref{eq:NHI}), eliminating $r_I$ using
Eq~(\ref{eq:rI}) and finally using Eq.~(\ref{eq:gCDDF}):
\begin{equation}
g(N_{\rm HI}, v_{90})
\equiv \frac{\partial^3 N}{\partial X\partial N_{\rm HI}\partial v_{90}}
= g(M_h, N_{\rm HI})\,\left(\frac{d v_{90}}{d\log M_h}\right)^{-1}\,.
\label{eq:gv90}
\end{equation}
Integrating this quantity over $N_{\rm HI}$, for $\log N_{\rm HI}[{\rm cm}^{-2}]\ge 20.3$, yields the line width distribution for \dla s. It turns out that not surprisingly most of the dependence of $v_{90}$ on halo mass arises from its dependence on the halo's virial velocity, so that approximately $d\log v_{90}/d\log M_h\approx d\log v_h/d\log M_h=1/3$ and hence $dv_{90}/d\log M_h\propto v_{90}$. Results are shown in Fig.~\ref{fig:v90}, where model results are shown at $z=3$.

In panel (a) of Fig.~\ref{fig:v90} we plot as {\em coloured lines} the relation between $v_{90}$ and the column density for various values of the halo mass; $M_h$, in units of ${\rm M}_\odot$, can be read from the legend. The more massive the halo, the higher $v_{90}$ at a given value of $N_{\rm HI}$, reflecting the $v_{90}\propto v_h$ dependence. The impact parameter decreases with increasing $N_{\rm HI}$ for a given value of $M_h$, as explained this increases $v_{90}$ as gas flows increasingly along the line of sight. For reference we show as {\em red symbols} the data of \cite{Neeleman13}. 
Panel (b) shows the fraction of \dla s with a given value of $v_{90}$ as a {\em black thick line}, as well as the line width distribution for two values of the column density, $\log N_{\rm HI}[{\rm cm}^{-2}]=20.3$ ({\em dark blue}) and 21.3 ({\em cyan}). This reveals as weak trend for lower column-density \dla s to have lower values of $v_{90}$. Shown as {\em red symbols} are the histogrammed data of \cite{Neeleman13}, showing that the model gives a reasonable distribution of velocity widths. 
In panel (c) we plot the number of lines with a given value of $v_{90}$ from Eq.~(\ref{eq:gv90}), for the same cuts in column-density as in panel (b). Over plotted as {\em red symbols} is the observed distribution for \dla s from \cite{Prochaska03}, which should be compared to the black line.

The model is in reasonable agreement with the data, producing a small number of quite broad lines with $v_{90}>400{\rm km~s}^{-1}$, and a large number of narrow lines with a rather abrupt cut-off at $v_{90}\lessapprox 25~{\rm km~s}^{-1}$, as observed. Comparing panels (b) and (c), it seems that the model slightly undershoots the number of narrow-lined \dla s when compared to the \cite{Prochaska03} data, whereas it overshoots the number of narrow-lined \dla s when compared to the \cite{Neeleman13} data. Either this means that the two data sets are not quite consistent, possibly just due to the small number statistics, or it may point to redshift evolution
in the data.

\subsection{Molecules}
\label{sect:molecules}
 \begin{figure*}
 \includegraphics[width=.95\textwidth]{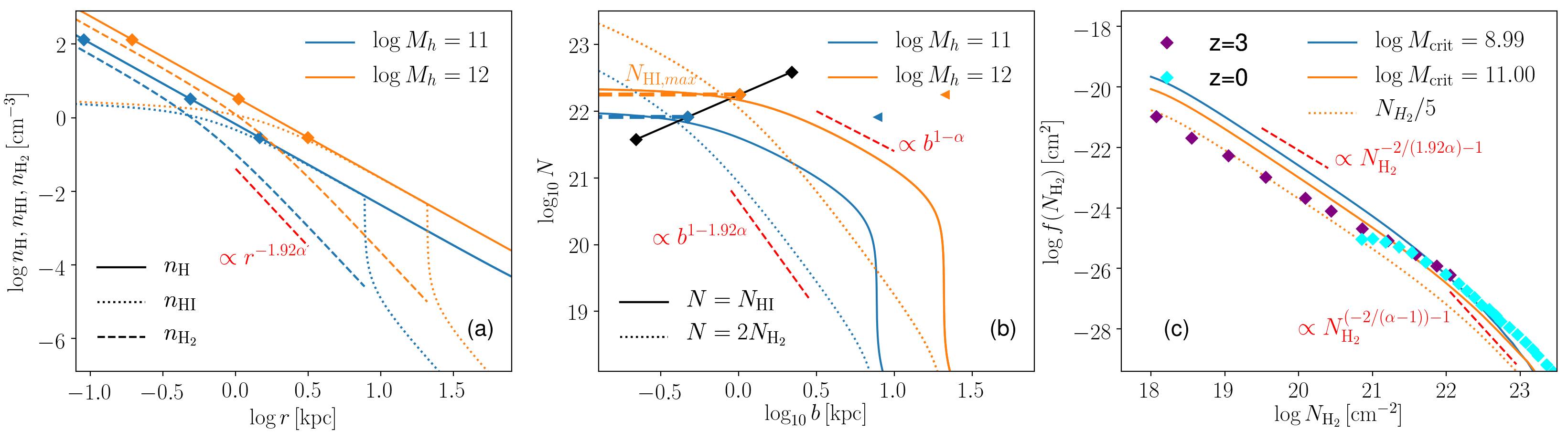}\hfill
 \caption{Molecular hydrogen in halos at redshift $z=3$. {\bf Panel (a):}
Assumed radial distribution of the 
total hydrogen density ($n_{\rm H}$, {\em solid line}), 
the atomic hydrogen density ($n_{\rm HI}$, {\em dotted line}) 
and the molecular hydrogen density ($n_{{\rm H}_2}$, {\em dashed line}) 
in halos with mass $M_h=10^{11}$ and $10^{12}{\rm M}_\odot$ ({\em blue and orange lines}, respectively); the power-law exponent is $\alpha=2.2$. The $n_{\rm H}$ and $n_{{\rm HI}}$ distributions are repeated from Fig.\ref{fig:radial}, except that now we subtract $2\times n_{{\rm H}_2}$ which causes the down-turn in the $n_{\rm HI}$ density at small radii. When the molecular fraction is low, $n_{{\rm H}_2}$ is approximately proportional to $n_{\rm H}^{1.92}$, with the factor 0.92 the exponent appearing in Eq.~(\ref{eq:BR}). This scaling is indicated by the {\em red dashed line}. At higher densities, the molecular fraction tends to unity so that $2n_{{\rm }H_2}\approx n_{\rm H}$. The left and right most {\em filled diamonds} correspond to the maximum and minimum density for which \protect\cite{Blitz06} measured the fitting relation
of Eq.~(\ref{eq:BR}); the middle {\em filled diamond} corresponds to the density where $R_{\rm mol}=1$. {\bf Panel (b):} Corresponding column-densities of atomic 
gas ({\em solid line}) and molecular gas ({\em dotted lines}) 
as a function of impact parameter, $b$. The slopes for pure power-law profiles are plotted as {\em red dashed lines}. When $N_{\rm HI}\gtrapprox 10^{22}{\rm cm}^{-2}$, gas in the model becomes increasingly molecular, with the transition column dependent on halo mass.
The {\em thick horizontal dashed lines} show the maximum {\sc HI} column-density
as computed from Eq.~(\ref{eq:NHmax}), with the {\em diamonds} at the location where $b=r_{{\rm H}_2}$, the radius within which $R_{\rm mol}>1$. The {\em black line} shows the relation $r_{{\rm H}_2}-N_{\rm HI, max}$, with further {\em black diamonds} indicating halo masses of $\log M_h[{\rm M}_\odot]=10$ and 13.
The {\em left pointing triangles} show the location of the ionization front, plotted at $b=R_I$.
{\bf Panel (c):} Corresponding molecular hydrogen column density distribution function. The {\em blue} and {\em orange} lines are the model prediction when summing over all halos more massive than $10^{8.99}$ and $10^{11}{\rm M}_\odot$, respectively. At column $N_{{\rm H}_2}\lessapprox
10^{22}{\rm cm}^{-2}$, gas is mostly atomic and the ${\rm H}_2$ {\sc cddf} is a power-law with slope $f(N_{{\rm H}_2})\propto N_{{\rm H}_2}^{-2/(1.92\alpha)}$, indicated by the upper {\em red-dashed line}. At higher column-densities, gas becomes mostly molecular, and the {\sc cddf} steepens to $f(N_{{\rm H}_2})\propto N_{{\rm H}_2}^{-2/(\alpha-1)}$ - the same slope as the
$N_{\rm HI}$ {\sc cddf} at high column-density -  indicated by the lower {\em red-dashed line}. 
These (approximate) power-law dependencies result from the dependence of cross-section on column-density (see text). The {\em purple diamonds} depict the observed {\sc cddf} for molecular hydrogen  at $z\approx 3$ reported by \protect\cite{Balashev18}, the {\em cyan crosses} are the $z=0$ data from \protect\cite{Zwaan06}. The {\em orange dotted line} assumes that the molecular abundances is 5 times lower than predicted by Eq.~(\ref{eq:BR}). 
}
 \label{fig:H2}
\end{figure*}

Since at least some fraction of \dla s correspond to a  sight line puncturing a galaxy, some \dla\ sight line should also intersect molecular gas. Here we try to estimate how often this occurs.
To do so we use the simple empirical pressure-${\rm H}_2/{\rm HI}$ relation from \cite{Blitz06}.

Measuring the surface density of local spiral galaxies in both neutral and molecular gas,
\cite{Blitz06} found that the ratio of these surface densities scaled with central gas pressure ($p$) approximately as (their equation 12)
\begin{equation}
R_{\rm mol}\equiv \frac{\Sigma({\rm H}_2)}{\Sigma({\rm HI})}=\left(\frac{p/k_{\rm B}}
{3.5\times 10^4\,{\rm K}{\rm cm}^{-3}}\right)^{0.92}\,,
\label{eq:BR}
\end{equation}
where $k_{\rm B}$ is Boltzmann's constant. In their paper, \cite{Blitz06} verify this relation in the range $3\times 10^3\le (p/k_{\rm B})/({\rm K}~{\rm cm}^{-3}) \le 2\times 10^6$. We will use this fit to relate the volume densities of
molecular and atomic hydrogen inside the ionization front, setting
\begin{equation}
\frac{2\,n_{{\rm H}_2}}{n_{\rm H}}\approx \left(1+\frac{1}{2R_{\rm mol}}\right)^{-1}\,.
\end{equation}
Given the assumed hydrogen density profile in halos, $n_{\rm H}\propto r^{-\alpha}$, and setting the gas temperature to $T=10^4{\rm K}$, we compute the gas pressure $p$, and substitute 
this value in the previous two relations to compute  the molecular hydrogen density, $n_{{\rm H}_2}$. Combining this with the number density of halos with a given mass allows us to compute the H$_2$ column-density distribution function. The results are illustrated in Fig.~\ref{fig:H2}.

The left panel repeats the (total) hydrogen profile, $n_{\rm H}\propto r^{-\alpha}$, with $\alpha=2.2$, from Eq.~(\ref{eq:profile}), replacing the atomic hydrogen density $n_{\rm HI}
\rightarrow n_{\rm HI}-2\,n_{{\rm H}_2}$ to account for molecular gas. The profiles are shown for two halo masses and apply to redshift $z=3$. The inclusion of molecules flattens the atomic profile towards the centre.  When $R_{\rm mol}\ll 1$ in self-shielded gas, $n_{{\rm H}_2}\approx R_{\rm mol}\,n_{\rm H}\approx n_{\rm H}^{1.92}$, since the gas pressure $p\propto n_{\rm H}$ when gas is isothermal. Therefore $n_{{\rm H}_2}\propto r^{-1.92\alpha}$ in the outskirts of the halo, shown in the panel as the {\em red-dashed} line.  At smaller radii, gas becomes mostly molecular so that $2n_{{\rm H}_2}\rightarrow n_{\rm H}$. The central diamond corresponds to the location where $R_{\rm mol}=1$, with the outer and inner diamonds enclosing the region to which \cite{Blitz06} fitted Eq.~(\ref{eq:BR}). As could have been anticipated, gas in the central regions is mostly molecular, in a shell around that it is mostly neutral, and in the outskirts it is highly ionized. Another point to take away from this is that there is a large range at low values of $n_{{\rm H}_2}$ in the neutral shell of gas where relation Eq.~(\ref{eq:BR}) is {\em extrapolated}: we should therefore treat the predicted values of the molecular gas fraction at these lower densities with caution.
 
The middle panel of  Fig.~\ref{fig:H2} shows the corresponding column-density as a function of impact parameter, $b$, computed as in Eq.~(\ref{eq:cdens}). As explained when discussing
Eq.~(\ref{eq:cdens_approx}), if the radial density profile is a power-law in radius, $\propto r^{-\beta}$, the column-density is approximately a power-law in impact parameter, $\propto b^{1-\beta}$. Using the values of $\beta$ from the left panel, we show the predicted power-law slopes as {\em dashed red lines}, for both the molecular and atomic column-densities. The power-law approximations fit the curves well over a large range in impact parameter. 

Comparing the $N_{\rm HI}$ profile with that in Fig.~\ref{fig:radial} shows that, not surprisingly,  including molecules decreases the ${\rm HI}$ column-density at small impact parameter, leading to a plateau in $N_{\rm HI}$ value at low $b$. The height of the plateau increases with halo mass, $M_h$. We can estimate the approximate value of this maximum column density at $b=0$ by assuming that gas in the spherical shell $r_{{\rm H}_2}\le r\le r_I$ is neutral, gas at $r> r_I$ is fully ionised, and gas at $r<r_{{\rm H}_2}$ is fully molecular. Here, $r_I$ is the radius of the ionization front from Eq.~(\ref{eq:rI}), and $r_{{\rm H}_2}$ is the radius within which the gas is mostly molecular. For the latter we use the value of the radius where $R_{\rm mol}=1$, as in the left panel. The maximum column-density is then
\begin{eqnarray}
N_{{\rm HI}, {\rm max} }&\approx& 2 \int_{r_{{\rm H}_2}}^{r_I}\,n_H(r)\,dr\nonumber\\
		&=&2\frac{n_{{\rm H},0}\,R_h}{\alpha-1}
	\left[
	\left(\frac{n_{{\rm H}_2}}{n_{{\rm H},0}}\right)^{(\alpha-1)/\alpha}-\left(\frac{R_h}{r_I}\right)^{\alpha-1}
	\right]\,,\nonumber\\
\label{eq:NHmax}
\end{eqnarray}
which indeed increases with increasing $R_h$ and hence $M_h$. The corresponding values are plotted in the middle of Fig.~\ref{fig:H2} as the {\em horizontal dashed lines}, for both values of $M_h$; they capture the numerical result quite accurately. The factor of 2 in Eq.~(\ref{eq:NHmax}) accounts for the sight-line crossing the neutral shell twice.

We also plot the values of $(b=r_{{\rm H}_2}, N_{{\rm HI}, {\rm max}})$ for halo masses $\log M_h[{\rm M}_\odot]=10$ and 13 in black, and for $\log M_h[{\rm M}_\odot]=11$ and 12 in blue and orange, respectively. For the last two values of $M_h$, we also plot the point $(b=r_I, N_{{\rm HI}, {\rm max}})$ as {\em left pointing triangles}. Comparing the triangles and diamonds, it is clear that for decreasing $b$, when $b\le r_I$, the {\sc HI} column density increases rapidly as gas becomes mostly neutral, and
once $b\le r_{{\rm H}_2}$, there is a kink in the {\sc HI} column-density-$b$ relation, because the gas interior to $r_{{\rm H}_2}$ is mostly molecular.

Given the run of the molecular column density as a function of $b$ and $M_h$, we now sum over all halos to compute the molecular hydrogen \cddf, shown in the right panel of Fig.~\ref{fig:H2}
at $z=3$. The {\em blue curve} sums the contribution of all halos with $M_h>M_{h, {\rm crit}}$ - the critical halo mass below which halos lose gas due to reionisation - as we did for the {\sc HI}\,\cddf. The slope of the \cddf\ is once more set by the dependence of the ${\rm H}_2$ cross-section on impact parameter, just as we found for the {\sc HI}\,\cddf. Given the power-laws from the middle panel, this predicts approximately scalings of 
$f(N_{{\rm H}_2})\propto N_{{\rm H}_2}^{(-2/(1.92\alpha)-1)}\approx N_{{\rm H}_2}^{-1.47}$ at low columns, $N_{{\rm H}_2}\lessapprox 10^{22}{\rm cm}^{-2}$, and
$\propto N_{{\rm H}_2}^{-2.66}$ above that. The steepening at high column-density is caused by the fact that these absorbers are mostly molecular, so that $2n_{{\rm H}_2}=n_{\rm H}$.
For comparison we over plot the observed \cddf\ at $z=3$ from \cite{Balashev18}, for which the slope at the lower column-densities is slightly flatter at $-1.29$. We also over plot the {\sc cddf} at $z=0$ from \cite{Zwaan06}.

There is no obvious reason for the $z=0$ data to smoothly fit onto the $z=3$ data - although, as \cite{Balashev18} point out, in fact they do. At high columns, $N_{{\rm H}_2}\gtrapprox 10^{22}{\rm cm}^{-2}$, our default model ({\em blue curve}) reproduces the $z=3$ data well, and fits smoothly onto the $z=0$ data. However below $N_{{\rm H}_2}\lessapprox 10^{21}{\rm cm}^{-2}$, the model increasingly overestimates the {\sc cddf}, although the observed and simulated slopes of the {\sc cddf}s are quite similar. Recalling our discussion of panel (a), we have extrapolated Eq.~(\ref{eq:BR}) to values much lower than measured in the paper by \cite{Blitz06}. It is these low densities and large impact parameters that
cause the overestimate in the {\sc cddf} at columns $N_{{\rm H}_2}\lessapprox 10^{21}{\rm cm}^{-2}$. We therefore suggest that Eq.~(\ref{eq:BR}) overestimates
$R_{\rm mol}$ at low pressures, possible as a consequence of the lower metallicity of this gas.
To illustrate how this would change the model's {\sc cddf}, we show with the {\em orange line} the default model, except we set the ${\rm H}_2$ abundance of halos with
$M_h<10^{11}{\rm M}_\odot$ to zero (to account for the reasonable expectation that lower mass halos have lower metallicity resulting in lower molecular fractions.). Such a change still leads to an over estimate of the {\sc cddf}. If we further reduce the ${\rm H}_2$ abundance of these halos by a  factor of five, we obtain the {\em orange dotted line}, which now agrees well with the data. From this comparison we conclude that, in so far as our model is actually applicable to molecules, low-density halo gas needs to be considerable less molecule rich than a naive extrapolation of Eq.~(\ref{eq:BR}) would suggest. A recent comparison of alternative models for molecule formation that include the effects of metallicity is presented by \citet{Schabe20}.

\subsection{$\Omega_{\rm HI}$, $\Omega_{{\rm H}_2}$ and their evolution}
\label{sect:OmegaHI}
\begin{figure*}
	\includegraphics[width=.95\textwidth]{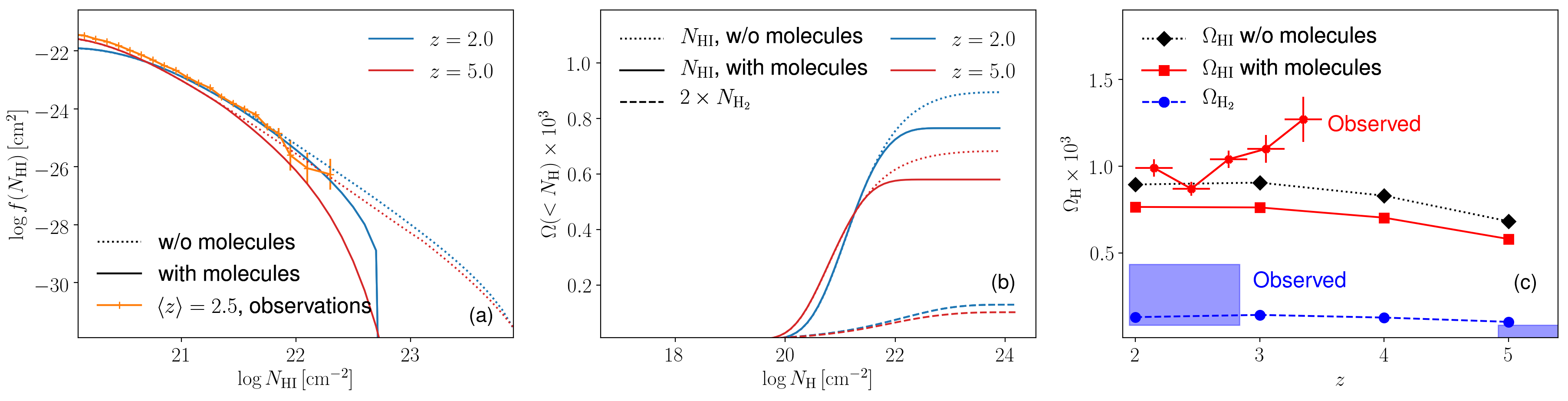}\hfill
	\caption{Density of neutral and molecular gas in units of the critical density. {\bf Panel (a)}: the model's {\rm HI} \cddf\ at two redshifts ({\em colours}) when molecules are neglected ({\em dotted line}) and when molecules are included ({\em solid line}). The {\em yellow line connecting error bars} is the observed \cddf\ from \protect\cite{Noterdaeme12}. {\bf Panel (b)}: as in panel (a), but plotting the cumulative fraction of gas in absorbers with column density $N_{\rm H}\leq N_{\rm HI}$.  The {\em dashed lines} correspond to the cumulative fraction of mass in molecular gas in absorbers with $N_{\rm H}\leq 2N_{{\rm H}_2}$.
	{\bf Panel (c)}: mass density in neutral gas in units of the critical density, Eq.~(\ref{eq:OmegaHI}), for the model when molecules are included ({\em red squares connected with a solid line}) and when molecules are not included ({\em black diamonds connected with a dotted line}). The {\em red line connecting error bars} are the observed values from \protect\cite{Noterdaeme12}. {\em Circles connected by a dashed line} refers to $\Omega_{{\rm H}_2}$, defined in Eq.~(\ref{eq:OmegaHI}); the {\em filled blue squares} are $\Omega_{{\rm H}_2}$ from the {\sc cold}z survey, \protect\cite{Riechers19}.}
	\label{fig:OmegaHI}
\end{figure*}
\begin{figure}
	\includegraphics[width=.95\columnwidth]{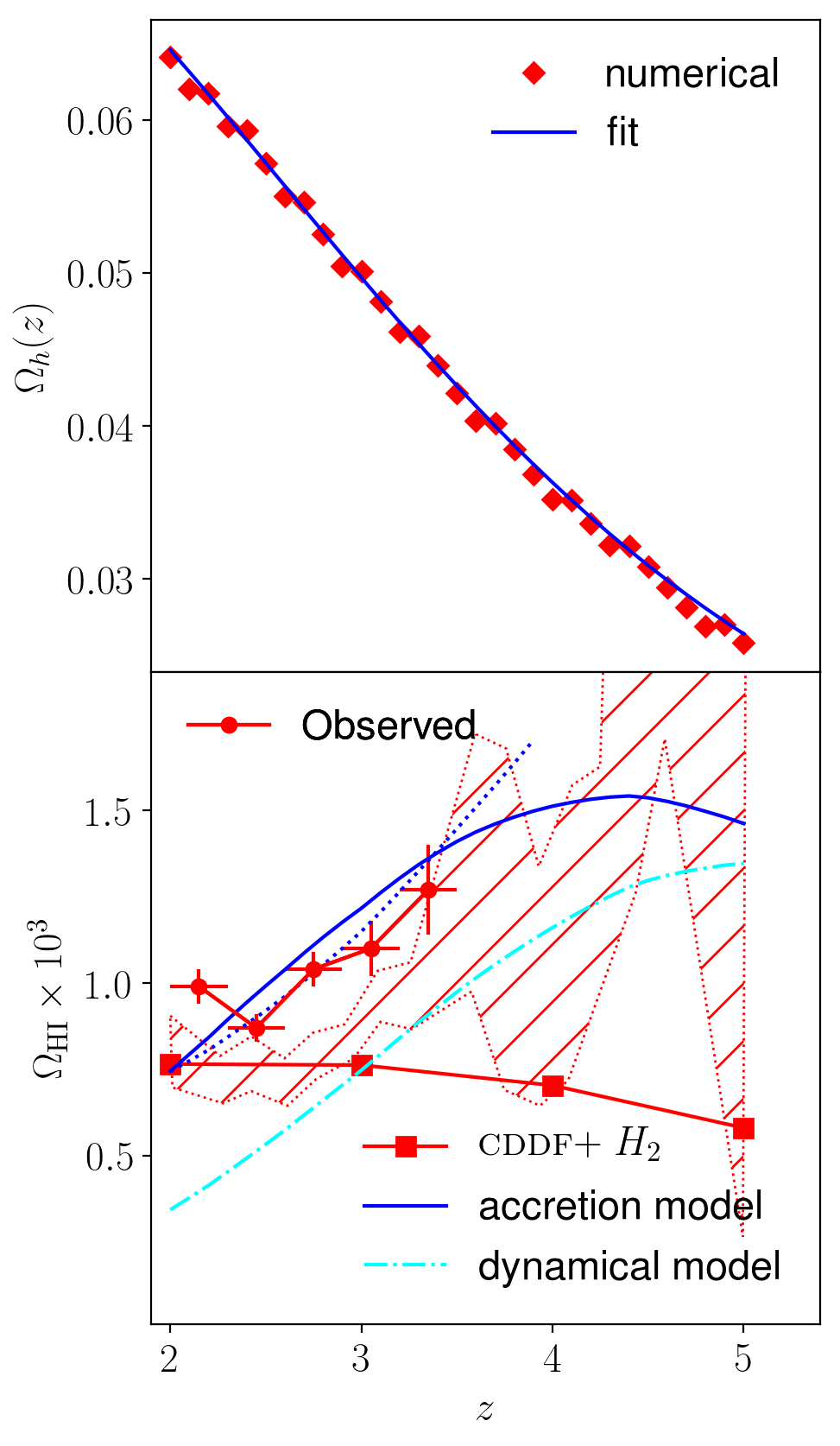}\hfill
	\caption{{\bf Upper panel:} Evolution of the collapsed fraction, Eq.~(\ref{eq:Omegah}), computed numerically ({\em red diamonds}) and the fit from Eq.~(\ref{eq:Omegah-fit}) ({\em blue line}). {\bf Lower panel:}
		Evolution of $\Omega_{\rm HI}$. The {\em red squares connected with a solid line} and {\em red line connecting error bars} are the model's evolution and the observed evolution, repeated from Fig.~\ref{fig:OmegaHI}, the {\em red hashed region} is the observed evolution from \protect\cite{Bird17}.  The {\em cyan dash-dotted line} and {\em blue solid line} 	assume that {\sc HI} gas is observable as it accretes onto halo for a fraction $\eta=r_I/R_h$ of the flow time, Eq.~(\ref{eq:OmegaHI}), using the dynamical time ($\tau=\tau_{\rm dyn}$, Eq.~\ref{eq:tau_d}) 
		or the accretion time ($\tau=\tau_{\rm acc}$, Eq.~\ref{eq:tau_a}), respectively. The {\em blue dotted line} is
		$\propto \eta(z)$, Eq.~(\ref{eq:eta_z}).}
	\label{fig:OmegaHI_z}
\end{figure}
Given the \cddf\ and the molecular fraction in halos, we can integrate over column density to compute the fraction of the cosmological mass that is in neutral and molecular gas in units of the critical density,
\begin{eqnarray}
\Omega_{\rm HI} &=&\frac{H_0\,\mu\,m_{\rm H}}{c\,\rho_c}\int_{0}^{\infty}N_{\rm HI}\,f(N_{\rm HI})\, dN_{\rm HI}\nonumber\\
\Omega_{{\rm H}_2} &=&\frac{H_0\,\mu\,m_{\rm H}}{c\,\rho_c}\int_{0}^{\infty} (2N_{{\rm H}_2})\,f(N_{{\rm H}_2})\, dN_{{\rm H}_2}\,.
\label{eq:OmegaHI}
\end{eqnarray}
Here, $\mu=1.3$ is the mean molecular weight per particle in units of the proton mass, $m_{\rm H}$, to account for Helium and other elements, the factor of two in the expression for $\Omega_{{\rm H}_2}$ counts the two hydrogen atoms per ${\rm H}_2$ molecule\footnote{Notice that this is the traditional definition of $\Omega_{\rm HI}$: the cosmological mean density in gas that is mostly neutral - as opposed to the mass density of {\sc HI} - in units of $\rho_c$.}. The predictions for these two quantities are illustrated in Fig.~\ref{fig:OmegaHI}.

Panel (a) of Fig.~\ref{fig:OmegaHI} repeats the {\sc HI} \cddf\ from Fig.~\ref{fig:cddf} at two redshifts ({\em dotted lines}), but also shows the \cddf\ corrected for molecules as discussed in the previous section ({\em solid lines}). The onset of ${{\rm H}_2}$ introduces a cut-off in the \cddf\ which is particularly sharp at $z=2$. Over most of the observed range ({\em yellow line connected error bars} is the data from \cite{Noterdaeme12}), the impact of including molecule formation is small \citep[see also][]{Schaye01}.

In panel (b) of Fig.~\ref{fig:OmegaHI} we plot the cumulative mass-density in neutral in units of $\rho_c$, $\Omega_{\rm HI}(<N_{\rm HI})$, at two redshifts ({\em different colours}) and with or without including ${{\rm H}_2}$ formation ({\em solid} and {\em dotted} lines, respectively). This panel demonstrates that $\Omega_{\rm HI}$ is dominated by gas in sub-\dla s and \dla s, as is well known. The panel also shows $\Omega_{{\rm H}_2}$ at the same two redshifts as {\em dashed lines}. There is little evolution in $\Omega_{{\rm H}_2}$ and
$\Omega_{{\rm H}_2}\ll \Omega_{\rm HI}$ - therefore including or excluding molecules makes less than $\sim 20$~per cent difference to $\Omega_{\rm HI}$.

In panel (c) of  Fig.~\ref{fig:OmegaHI} we plot the evolution of $\Omega_{\rm HI}$ ({\em solid} and {\em dotted} lines show the model including or excluding molecules, respectively), compared to the data of \cite{Noterdaeme12}. The model prediction is within 20~per cent of the observed value. Within the observational error bars, there is little evidence for evolution in the data as has been stressed by e.g. \citet{Prochaska09}.
That panel also compares the predicted value of $\Omega_{{\rm H}_2}$ ({\em blue circles connected by a dashed line}) to the value reported by \cite{Riechers19} from the {\sc cold}z survey (the {\em blue filled rectangles}, which correspond to the green error boxes in Figure~5 of \citealt{Riechers19}). Given the relative simplicity of the model, the prediction is in relatively good agreement with the data. There is a hint that the evolution in the data is stronger than in the model.

In models in which {\sc HI} is mostly associated with the {\sc ism} of galaxies, the non-evolution of $\Omega_{\rm HI}$ over the redshift range $z=5\rightarrow 2$  is surprising because the galaxy stellar mass function builds-up significantly over this interval. If {\sc HI} is the gas reservoir associated with galaxies, then we might expect $\Omega_{\rm HI}$ and $\Omega_\star$ (the cosmological mass density in stars divided by $\rho_c$) to evolve in tandem - but they do not, see also \citet{Peroux12}. However, in the current model, most of the {\sc HI} mass is actively accreting onto halos rather than a reservoir that is being build-up. As gas accretes onto a  halo, it is initially highly ionised and hence does not contribute to $\Omega_{\rm HI}$. Once gas flows past the ionization front, $r<r_I$ from Eq.~(\ref{eq:rI}), it becomes neutral. Even further in, $r<r_{{\rm H}_2}$, the radius where $R_{\rm mol}$ from Eq.~(\ref{eq:BR}) becomes $\sim 1$, gas becomes molecular. This means that the cosmological mean density of {\sc HI}, $\rho_{\rm HI}=\Omega_{\rm H I}\,\rho_c/\mu$, is due to gas in halos at a distance from the centre $r_{{\rm H}_2}<r<r_I$. We can compute the evolution of the cosmological density of such gas as follows.

We first notice from panel (b) in Fig.~\ref{fig:OmegaHI} that $\Omega_{\rm HI}$ is dominated by gas with neutral fraction $x\approx 1$. This means that its numerical value does not depend (strongly) on the radial distribution of gas in halos. Indeed, if instead $\Omega_{\rm HI}$ were to be dominated by the small neutral fraction in highly ionised gas for which $n_{\rm HI}\propto n_{\rm H}^2$ according to Eq.~(\ref{eq:xprofile}), then $\Omega_{\rm HI}$ would depend strongly on the radial density profile, $n_{\rm H}(r)$. Since we concluded that $\Omega_{\rm HI}$ does not depend on the radial distribution (as long as molecules can be neglected as well),  $\rho_{\rm HI}$ should be related to the rate at which mass collapses into halos that contain \dla s, {\em i.e.} those with halo mass $M_h>M_{\rm h, crit}$. This means that we should be able to compute  $\Omega_{\rm HI}$ directly from the rate of collapse into halos without computing the \cddf\ first. We do so in the following.

We compute the collapsed fraction in halos that host \dla s as the integral 
\begin{equation}
\Omega_h \equiv \frac{1}{\rho_c}\int_{\log M_{\rm h, crit}}^\infty M_h\,\frac{dn}{d\log M_h}\,d\log M_h\,,
\label{eq:Omegah}
\end{equation}
where in practise we integrate to an upper limit of $M_h=10^{13}{\rm M}_\odot$. This does not affect the results here, since the halo mass function is already decreasing exponentially well before that. We evaluate the integral in Eq.~(\ref{eq:Omegah}) numerically, computing $dn/d\log M_h$ with the \col\ tool of \cite{Diemer18}.
The result is plotted in the top panel of Fig.~\ref{fig:OmegaHI_z}, together with the Taylor-series fit
\begin{equation}
\ln\left(\frac{\Omega_h(z)}{0.065}\right)=-0.22(z-2)-0.05(z-2)^2+0.0008(z-2)^3\,,
\label{eq:Omegah-fit}
\end{equation}
from which we compute that
\begin{equation}
\frac{d\Omega_h}{dz}\approx -0.015 (1 + 0.23\,(z-2) - 0.18\,(z-2)^2+\cdots)\,,
\end{equation}
and $d\Omega_h/dz \approx -0.015$ is constant to within $\sim 20$~per cent from $z=2\rightarrow 4$.

Our reasoning for the computation of $\Omega_{\rm HI}$, given $d\Omega_h/dt=\dot z\,d\Omega_h/dz$, goes as follows. Gas accretes onto halos that host \dla s at a rate $\omega_b\,d\Omega_h/dt$, where the factor $\omega_b\equiv \Omega_b/\Omega_m$ accounts for the gas fraction of the accreting matter. This accreting gas is observable as {\sc HI} over a fraction, $\eta$, of the time $\tau$ it takes to flow from the the virial radius of the halo, $r=R_h$ to the centre, $r=0$. We can estimate $\tau$ in two ways, which should give approximately the same answer. We can set $\tau=\tau_{\rm dyn}$, the dynamical time of the halo, or we can set $\tau=\tau_{\rm acc}\equiv R_h/v_r$, the \lq accretion time\rq\ of the halo, with the radial velocity of the gas taken from Eq.~(\ref{eq:vr}).

The dynamical time of the halo, $\tau_{\rm dyn}$, is approximately
\begin{equation}
\tau_{\rm dyn} = \frac{1}{({\rm G}\bar\rho_h)^{1/2}}=\left(\frac{8\pi}{600}\right)^{1/2}\frac{1}{H(z)}\,,
\label{eq:tau_d}
\end{equation}
where the second step sets the mean density $\bar\rho_h$ within $R_h$ to 200 times the critical density \cite[e.g.][]{Mo98}. The accretion time of the halo, $\tau_{\rm acc}$, is
\begin{eqnarray}
\tau_{\rm acc}&\equiv &\frac{R_h}{v_r}=\left(\frac{d\ln m_h}{dz}\dot z\right)^{-1}\
                      \approx \frac{1.33}{(1+z)H(z)}\,,
\label{eq:tau_a}                      
\end{eqnarray}
where we used Eq.~(\ref{eq:correaM}) for the halo growth rate taking the numerical value $b=0.75$ from \cite{Correa15a}. Our result is approximate in that we have set $\alpha=2$ to calculate the halo's gas mass. Finally, the fraction of time that the gas is observable as {\sc HI} should be approximately\footnote{
For an $1/r^2$ density distribution - which is close to the radial profile that we assume when taking $\alpha=2.2$ - the enclosed mass in a {\em sphere} of radius $r$ is proportional to $r$. However, here we need the mass enclosed in a circular aperture. Since we take a single value of $\eta$, whereas in reality we should take the average, weighing each halo mass with its cross-section,  we think  this approximation is good enough to understand the scaling of $\Omega_{\rm HI}$.} $\eta\approx r_I/R_h$, the ratio of the radius of the ionization front ($r_I$) - within which the gas is neutral - over the viral radius ($R_h$).

Given this reasoning, we compute $\Omega_{\rm HI}$ as
\begin{equation}
\Omega_{\rm HI} = \left(\omega_b\,\frac{d\Omega_h}{dt}\right)\,\times (\eta\,\tau)\,,
\label{eq:OmegaHI}
\end{equation} 
where $\tau=\tau_{\rm dyn}$ or $\tau=\tau_{\rm acc}$. 

This equation forecasts that $\Omega_{\rm HI}$ depends relatively weakly on redshift. Indeed, taking $\tau=\tau_{\rm acc}$ we find that $\Omega_{\rm HI}(z)\propto \eta(z)\,(-d\Omega_h/dz)$. Given that 
$d\Omega_h/dz$ is approximately constant, the redshift dependence of $\Omega_{\rm HI}$ is mostly due to that of $\eta(z)$, which we discuss in more detail below. The reason for the lack of evolution is illuminating: 
it occurs because the rate of collapse into halos is approximately inversely proportional to the accretion time of those halos. Therefore the product, $(d\Omega_h/dt)\,\tau_{\rm acc}$, is almost constant. This is the main reason for the lack of evolution in $\Omega_{\rm HI}$ in the current model.

We examine the evolution of $\eta$ starting from Eq.~(\ref{eq:rI}), 
\begin{eqnarray}
\eta(z) &\approx& 0.25\,\left(\frac{ (n_{H,0}(z)/n_{\rm H,0}(2))^2\,(R_h(z)/R_h(2))}{\Gamma(z)/\Gamma(2)}\right)^{1/(2\alpha-1)}\nonumber\\
&\approx& 0.25 \frac{((1+z)/3)^{5/(2\alpha-1)}}{(\Gamma(z)/\Gamma(2))^{1/(2\alpha-1)}}\,.
\label{eq:eta_z}
\end{eqnarray}
We used $z=2$ as the reference point, reading the value of $\eta(z=2)\approx 0.25$ for a halo with mass $M_h=10^{11}{\rm M}_\odot$ from Fig.~\ref{fig:radial}. The virial radius of a halo with given mass decreases with $z$ as $R_h\propto 1/(1+z)$, whereas the density of the halo increases $\propto (1+z)^3$. As a result $r_I$ {\em increases} whereas $R_h$ decreases with $z$. This evolution, combined with the evolution of $\Gamma$, causes the ratio $r_I/R_h$ to increase with $z$.

The numerical values we obtain are
\begin{equation}
\Omega_{\rm HI}(z=2)=
	\begin{cases}
	0.74\times 10^{-3}\quad \hbox{for $\tau=\tau_{\rm acc}$,}\\
	0.34\times 10^{-3}\quad \hbox{for $\tau=\tau_{\rm dyn}$}\,.
	\end{cases}
\label{eq:accretion}
\end{equation}
The predicted evolution for both models is plotted in the lower panel of Fig.~\ref{fig:OmegaHI_z}
({\em solid line}: accretion model which uses $\tau=\tau_{\rm acc}$, {\em dash-dot line}: dynamical model which uses $\tau=\tau_{\rm dyn}$). The value obtained for the accretion model ($\Omega_{\sc HI}(z=2)=0.75\times 10^{-3}$) agrees very well with that obtained by integrating the $z=2$ \cddf\ ($\Omega_{\sc HI}(z=2)=0.77\times 10^{-3}$) for the full model ({\em solid squares connected by a red line}) as well as the value inferred observationally by \cite{Noterdaeme12} ({\em solid red line connecting error bars}, ($\Omega_{\sc HI}(z=2.2)=(0.99\pm 0.05)\times 10^{-3}$) repeated from Fig.~\ref{fig:OmegaHI} in the lower panel of Fig.~\ref{fig:OmegaHI_z}. The dynamical model under estimates 
$\Omega_{\rm HI}$ by a factor of $\sim 2$.

We stress that these models are meant to reproduce the full model, shown as {\em red squares connected with a solid line}. The stronger evolution in the accretion model compared to the full model is due to the stronger evolution in $\eta(z)$ in the former. This is demonstrated by the {\em blue dotted line}, which is $\propto \eta(z)$ - it captures the evolution of the accretion model well. What the accretion model does not capture is that, at higher $z$, the \cddf\ and hence $\Omega_{\rm HI}$, is dominated by lower-mass halos. Such lower-mass halos have lower values of $\eta$. The accretion model does not account for this evolution of the \dla\ halo mass function. 
%
%
%
%
%
%

In summary: the value of $\Omega_{\rm HI}$ obtained by integrating the model \cddf\ agrees well with the value observed by \cite{Noterdaeme12}. Neither model nor data evolve strongly with redshift. In the model, {\sc HI} gas is accreting onto halos. We discussed how such an accretion model can be derived directly without making reference to the \cddf, Eq.~(\ref{eq:OmegaHI}). In this accretion model, the relative constancy of
$\Omega_{\rm HI}$ results mostly from the fact that the accretion rate scales $\propto H(z)$ whereas the dynamical time of a halo scales $\propto 1/H(z)$ so that $\Omega_{\rm HI}$, which is proportional to their product, depends weakly on $z$.

\section{Discussion}
\label{sect:discussion}
We have made several simplifications in developing the simple model for predicting the absorption properties of gas as it accretes onto halos and feeds a central galaxy. We briefly discuss their impact and how they could be improved.

\noindent$\bullet$ {\em Spherical accretion.} Numerical simulations show convincingly that cold gas accretes onto halos in the form of {\em filaments} rather than spherically symmetric
\citep[e.g.][]{Fumagalli11, Cen12, Bird14, Rahmati15, Vandevoort19}. Observational evidence for such filamentary accretion is mounting \citep[e.g][]{Fumagalli17}. At the very least, the accreting gas also has angular momentum which will affect how it accretes. In addition, galactic outflows driven by energy injected in the interstellar medium are a major ingredient in simulations of galaxy formation (see e.g \citealt{Somerville15} for a review). Such outflows must, at some level, impact the accretion of gas. Finally, as gas accretes onto the galaxy, it eventually must encounter an accretion shock. Assuming that gas accretion is spherical, as we have done in this paper, is clearly a major simplification. One reason we suggest that this may not be as unreasonable as it looks at first sight, is that a major fraction of the \dla\ cross section of gas in a halo occurs relative far out. That accreting gas may not be affected so much by angular momentum or bipolar galactic outflows. In addition, provided that the halo mass is low, the location of any accretion shock may be relatively close to the galaxy, in which case it does not strongly affect the \dla\ cross section. All these processes are likely to affect the \cddf\ at higher values of the column-density. There is support for this view from the simulations by \cite{Altay11}, who show that the \cddf\ is little affected by feedback for columns $N_{\rm HI}\le 10^{21.5}{\rm cm}^{-2}$, but there is increasingly significant impact at higher column-densities.

\noindent $\bullet$ {\em {\sc HI} in the central galaxy.} At the centre of the halo, some fraction of the accreted gas will remain in the form of {\sc HI} in the galaxy, with star formation occurring in molecular gas.
We have not accounted for such a \lq reservoir\rq\ of gas, but did try to account for ${\rm H}_2$ formation.
For the model to be viable, the majority of the accreted gas must therefore be ejected again in the form of a galactic wind, rather than be contained in the galaxy, and moreover this wind should not itself contributes significantly to {\sc HI} in absorption. The galaxy formation model by \cite{Sharma20} makes it plausible that
the galactic outflow rate traces the accretion rate, rather than building a gas reservoir in the galaxy.
Most models of galaxy formation appeal to strong outflows, see e.g. \cite{Somerville15} for a recent review.

\noindent $\bullet$ {\em Cosmological accretion} We assumed that the matter that accretes onto halos does so purely in the form of cosmological accretion with the cosmic baryon fraction. In reality, some fraction of baryons will have collapsed into smaller galaxies before, and some may have been ejected by feedback. Conversely, a fraction of the gas accreting onto the halo may have been ejected by the halo's galaxy previously and is now being \lq recycled\rq. The extent to which this affects accretion onto halos likely depends on mass and redshift,  \cite{Garratt20, Wright20} examined some of these effects in the {\sc eagle} simulations at $z\lessapprox 2$.

\noindent  $\bullet $ {\em Recombinations.} We used the \lq case-B\rq\ recombination rate for hydrogen, appropriate in situations where a recombination directly to the ground state releases a photon that ionizes a neutral hydrogen in the immediate vicinity - also called the \lq on-the-spot approximation\rq. This might be a good approximation inside and close to the ionization front, but will underestimate recombinations in the highly-ionized outskirts of the gas profile. At $T=10^4{\rm K}$, the case-A (total) recombination rate is $\approx 1.6$ times higher than the case-B value.  Improving this aspect might increase the number of LLS and sub-\dla s by a small fraction, and could also affect the shape of the knee in the \cddf. In a similar vein we also neglected spectral hardening, which could also impact the shape of the knee in the \cddf. Both effects were included in the {\sc urchin} radiative transfer calculations of \cite{Altay11} to which we compared our model in Fig.~\ref{fig:cddf_evolution}.  In particular, {\sc urchin} switches between case-A and case-B recombination rates when the shielding optical depth $\tau>1$ \citep{Altay13b}. It also uses 100 frequency bins to account for spectral hardening. We also assumed that the accreting gas remains isothermal at $T=10^4~{\rm K}$. The temperature-density relation seen in cosmological simulations shows that {\sc igm} gas instead heats as gas accretes from $T\sim 10^4{\rm K}$ to $\approx 10^{4.4}~{\rm K}$ before cooling back to $10^{4}~{\rm K}$ (see for example Fig.~2 in \citet{Theuns98}). This increase in $T$ decreases the recombination rate by a factor $\approx 2.2$. To get this temperature evolution right requires radiation-hydrodynamical simulations, which can simultaneously account for shielding from the UV-background and its impact on the cooling and heating rate of the accreting gas. We also neglected any impact of radiation from the central galaxies on the \dla\ \citep[see e.g.][]{Rahmati13}.

\noindent $\bullet $ {\em \dla\ line-widths.} A characteristic feature of the line-shape of low-ionization metal absorption associated with \dla s is that the strongest absorption tends to occur at the edges of the absorption systems (\lq edge-leading spectra\rq, \citealt{Neeleman13}). Simulations seem to be able to reproduce this \cite[e.g.][]{Bird15}. We have not examined whether our model is consistent with this.

\noindent$\bullet${\em Molecules.} At face-value, the agreement between the model prediction and observations of the molecular \cddf\ is not very good, with the model over predicting $N_{{\rm H}_2}$ by a significant amount. It would be interesting to examine how well simulations do. It seems reasonable to expect that a model that accounts better for the impact of metals on the ${\rm H}_2$ abundance would improve the agreement with the data. In any case, we expect that some model features would still emerge, such as the flattening of the $N_{\rm HI}$ column at small impact parameter parameter and the associated bend in the ${\rm HI}$ \cddf\ at high $N_{\rm HI}$.

\section{Summary and conclusions}
\label{sect:conclusions}
The high star formation rate measured in galaxies at $z\gtrapprox 2$ is mostly fuelled by cosmological accretion, however observing this accreting gas directly has proved to be challenging \cite[e.g.][]{Fumagalli11}. In this paper we have examined a simple model for cosmological accretion onto dark matter halos, based on the similarity solution of \cite{Bertschinger85}, and including approximate radiative transfer to account for self-shielding from an ambient ionising background taken from \cite{Haardt12}. In this model, the accreting gas gives rise to Lyman-limit systems (LLS) in the surroundings and outskirts of the halo where it is highly ionised, and to damped Lyman-$\alpha$ systems (\dla s) in the inner halo where the gas is mostly neutral. The resulting column-density distribution function (\cddf) is in excellent agreement with the data (Fig.\ref{fig:cddf}). To the extent that the model captures accretion of gas onto halos, it seems we have been observing the accreting gas
that fuels star formation all along.

The shape of the \cddf\ reflects that of the power-law radial distribution of the gas, $n_{\rm H}(r)\propto r^{-\alpha}$, with $\alpha\approx 2.2$. The reason is that the density profile is self-similar and hence the same for all halos.  In the LLS regime, the slope of the \cddf\ is $f(N_{\rm HI})\propto N_{\rm HI}^{2/(1-2\alpha)-1}\propto N_{\rm HI}^{-1.6}$, whereas in the DLA regime it is
$f(N_{\rm HI})\propto N_{\rm HI}^{2/(1-\alpha)-1}\propto N_{\rm HI}^{-2.7}$. Our analytical expression explain why halos contribute about equally to the \cddf\ over a large range of halo masses, $M_h=10^{10}-10^{12}{\rm M}_\odot$ (lower-mass halos are more numerous but their cross-section is smaller, see Eq.~\ref{eq:gz}) as well as why the amplitude of the \cddf\ evolves so little with redshift (although halos become more abundant at lower $z$, their cross-section decreases, see Eq.~\ref{eq:z-scaling}). We also explain the origin of a lower-mass cutoff (reionization introduces a redshift-depend critical mass, $M_{\rm crit}(z)$,  below which halos lose their gas, \cite{Okamoto08}) and why high-mass halos do not contribute more to \dla s (the exponential cut-off in the halo mass function).

The relative contribution of halos of a given mass to the \cddf\ does evolve in the sub-\dla\ regime 
($N_{\rm HI}\lesssim 10^{20}{\rm cm}^{-2}$), where lower-mass halo contribute more at higher $z$. This leads to a change in the location of the \lq knee\rq\ in the \cddf\ - the transition from LLS to \dla s- with the knee shifting to lower values of $N_{\rm HI}$ at higher $z$ (see Fig.~\ref{fig:cddf}). This shift is caused by the evolution of the critical mass $M_{\rm crit}(z)$, which becomes smaller at higher $z$. Detecting such an evolution in the data might constrain the evolution of this critical mass.

Computing the differential contribution of halos of a given mass to the \cddf\ allows is to evaluate the \dla\ bias (\S \ref{sect:bias}) , which agrees well with the observed value, as well as the distribution of 
\dla\ line-widths, $v_{90}$ (\S \ref{sect:v90}). We assumed that accreting gas falls in radially, at a rate consistent with the cosmological accretion rate, and weigh the infall velocity with the cosine of the angle with the line of sight. For a {\em given} halo, this introduces a relation between column-density and $v_{90}$, because at a smaller impact parameter, the \dla\ column density is higher and the gas flows increasingly parallel to the line of sight so that $v_{90}$ is larger. Integrating over all halos yields the distribution of $v_{90}$, which has a relatively sharp cut-off at $\sim 30~{\rm km~s}^{-1}$ and a long tail towards high values of $v_{90}$
(Fig.\ref{fig:v90}). The sharp cut-off is related to the value of $M_{\rm crit}(z)$. The extended tail results from \dla\ sight-lines that intersect rare massive halos at an unusually small impact parameter.

We used the model of \cite{Blitz06} to attempt to compute where the \dla\ gas turns molecular (\S \ref{sect:molecules}). This over predicts the ${\rm H}_2$ \cddf, especially at low $N_{{\rm H}_2}$, but does reasonably well at higher column-densities. The model predicts a maximum value of $N_{\rm HI}$ caused by high column-density gas turning increasingly molecular \citep[see also][]{Schaye01, Erkal12}. The value of this maximum column-density depends on halo mass. The presence of such a maximum leads to a down-turn in the {\sc HI}\,\cddf. Notwithstanding the superb statistics of the observed \cddf\ \citep{Noterdaeme12}, such a down-turn is not yet clearly detected.

Integrating over the \cddf\ yields the fraction of the mass in the Universe that is in neutral gas,
$\Omega_{\rm HI}$, or in molecular gas, $\Omega_{{\rm H}_2}$  (\S \ref{sect:OmegaHI}). 
The model prediction agrees well with observations (Fig.\ref{fig:OmegaHI}). We presented a model that captures the accretion origin of the {\sc HI} gas, and which reproduces well the results of the full model. This simpler
model elucidates why $\Omega_{\rm HI}$ evolves so slowly,  as observed. The underlying reason is that the accretion rate onto halos increases with redshift at the same rate as the flow time scale of the gas within halos decreases. Since $\Omega_{\rm HI}$ is proportional to the product of these (Eq.~\ref{eq:OmegaHI}),
it evolves slowly.

It has long be posited that the {\sc HI} seen in absorption fuels star formation in galaxies \citep[e.g][]{Wolfe05}. Within this interpretation, the fact that the mean {\sc HI} density, $\Omega_{\rm HI}$, evolves weakly with $z$ whereas the star formation rate density evolves strongly, is puzzling. The solution to the puzzle as discussed in this paper is twofold: firstly, the observed {\sc HI} is dominated by gas accreting onto halos rather than a reservoir of gas in the galaxy's interstellar medium (ISM). Although a sight-line through a galaxy likely produces a \dla, the reverse is not true: {\em every galaxy is a \dla, but not every \dla\ is a galaxy}. The accreting gas is in the form of {\sc HI} for a fraction of time as it accretes onto a halo. This explains why the accretion rate (and hence also the star formation rate) can vary with redshift while  $\Omega_{\rm HI}$ remains approximately constant. Secondly, as the {\sc HI} gas gets close to the centre and enters the galaxy's ISM, it becomes molecular, some fraction forms stars, but the majority is ejected in the form of a galaxy-wide outflow. This implies that any \lq reservoir\rq\ of {\sc HI} in the ISM of the galaxy remains small.

In short we claim that Lyman-limit systems (LLS) and \dla s are both related to the accretion of gas onto halos, and only a small fraction of this accreted gas fuels star formation in galaxies. This simple picture is consistent with the observed properties of LLS and \dla s.

\section*{Acknowledgements}
I want to thank J Fynbo for encouraging me to start this project, M Fumagalli and R Cooke for excellent suggestions, and an anonymous referee for constructive comments. This work was supported by STFC grant ST/P000541/1 and used the DiRAC Data Centric system at Durham University, operated by the Institute for Computational Cosmology on behalf of the STFC DiRAC HPC Facility (www.dirac.ac.uk). This equipment was funded by BIS National E-infrastructure capital grant ST/K00042X/1, STFC capital grants ST/H008519/1 and ST/K00087X/1, STFC DiRAC Operations grant ST/K003267/1 and Durham University. DiRAC is part of the UK's National E-Infrastructure.
\section*{Data Availability}
No new data were generated or analysed in support of this research.
\bibliographystyle{mnras}
\bibliography{theorybib}
\end{document}